\documentclass[twocolumn,final,journal,twoside]{IEEEtran}
\usepackage{amsmath}
\usepackage{latexsym}
\usepackage{amssymb}
\usepackage{bm}
\usepackage[dvips]{graphicx}

\newtheorem{defi}{Definition}
\newtheorem{lem}[defi]{Lemma}
\newtheorem{thm}[defi]{Theorem}
\newtheorem{rem}{Remark}

\def\Tr{\mathop{\rm Tr}\nolimits}

\def\argmax{\mathop{\rm argmax}\nolimits}
\def\real{\mathbb{R}}

\def\Label#1{\label{#1}\ [\ #1\ ]\ }
\def\Label{\label}

\begin{document}
\title{General formulas for\\
fixed-length quantum entanglement concentration}
\author{
Masahito Hayashi
\thanks{M. Hayashi was with Laboratory for Mathematical Neuroscience, 
Brain Science Institute, RIKEN.
2-1 Hirosawa, Wako, Saitama, 351-0198, Japan.
Now, he is with 
Quantum Computation and Information Project,
Japan Science and Technology Agency.
201 Daini Hongo White Bldg.
5-28-3, Hongo, Bunkyo-ku, Tokyo 113-0033, Japan.
(e-mail masahito@qci.jst.go.jp)}}
\markboth{}{HAYASHI: General formulas for
fixed-length quantum entanglement concentration}
\date{}
\maketitle

\begin{abstract}
General formulas of entanglement concentration
are derived by using an information-spectrum approach
for the i.i.d.\ sequences and 
the general sequences of partially entangled pure states.
That is,
we derive general relations
between the performance of the entanglement concentration
and the eigenvalues of the partially traced state.
The achievable rates with constant constraints and
those with exponential constraints
can be calculated from these formulas.
\end{abstract}


\section{Introduction}\Label{s1}
\PARstart{V}{arious} quantum information processings
are proposed, 
many of which
require maximally entangled states as 
resources, {\it e.g.}, 
quantum teleportation and dense coding {\it etc}\cite{B2,B1,ekert}.
Hence, it is often desired to generate maximally entangled states.
However, the realized state is not necessarily 
a maximally entangled state.
Thus,
entanglement concentration is used for producing maximally 
entangled states (MES) from partially entangled pure states
only by local operation and classical communication (LOCC),
while entanglement distillation is used for
producing them from partially entangled mixed states by LOCC.
Therefore, entanglement concentration
is an important issue in the field of quantum information.

In information theory, we often assume that
the system is prepared as 
the independent and identical multiple copies of the given state.
Such a condition is called independently and identically
distributed (i.i.d.) condition.
Under this condition,
Bennett {\it et al}.\cite{Ben} showed that
the amount of entanglement of 
a partially entangled pure state $|\Phi \rangle \langle \Phi|$
is described by the entropy $H(\rho)$ of its partial traced state
$\rho := \Tr_{{\cal H}_B}|\Phi \rangle \langle \Phi|$,
which is called the reduced density matrix.
That is, they proved that
an MES with size $2^{n H(\rho)}$ can be asymptotically produced
from $n$ identical copies of the state $|\Phi \rangle \langle \Phi|$
with a high enough probability.
Furthermore, 
independently of the form of $|\Phi \rangle \langle \Phi|$,
Hayashi and Matsumoto constructed 
a protocol satisfying the above property,
which is called universal \cite{HM}.

However, in the correlated physical system,
the state of the total system cannot be regarded as
independent and identical copies of a given state.
In such a case, we have to treat general partial entangled pure state 
between two distinct parties.
Indeed, this model is not so unnatural
because the state on the total system is pure when 
this system is isolated from the other system.
In this paper, as a general asymptotic method to treat 
this model asymptotically, 
we focus on the information spectrum method and apply it to
entanglement concentration.
The information spectrum method 
has been developed by
Han and Verd\'{u} \cite{HanVerdu_output}
for discussing general sequence of information sources/channels,
and been established as 
a unified method to information theory in Han's textbook\cite{Han_book}.
Indeed, this method has been applied to quantum information theory,
for example, to quantum hypothesis testing\cite{NH} and
quantum channel coding\cite{HN}.
In this paper, 
we apply this method to entanglement concentration,
and characterize the asymptotic production rate 
of a general sequence of partially entangled pure states
without any assumption.
The information spectrum method used in this paper
is slightly different from the original Han-Verd\'{u}'s method,
and is close to Nagaoka-Hayashi's method\cite{NH}.

In the derivation of our general asymptotic formulas, 
we essentially use the majorization method 
established by Nielsen\cite{N}.
Based on this method,
he developed a necessary 
and sufficient condition for the possibility
of transforming from 
a partially entangled pure state $|\Phi_1 \rangle \langle \Phi_1|$
to another entangled pure state $|\Phi_2 \rangle \langle \Phi_2|$ 
by using LOCC between the two parties
${\cal H}_A$ and ${\cal H}_B$.
This condition is characterized
only by the eigenvalues of their
reduced densities $\rho_i:= 
\Tr_{{\cal H}_B} |\Phi_i \rangle \langle \Phi_i|,~(i=1,2)$.

Moreover, even in the i.i.d.\ case, 
the knowledge of the asymptotic production rate is not sufficient 
for estimating the production rate of MES 
for a given finite number of copies.
In channel coding or source coding,
for this analysis, we usually focus on the error exponents,
{\it i.e.}, the exponential rate of error probability
because the error goes to $0$ exponentially when we choose our code suitably.
In entanglement concentration,
when we fix the production rate to a constant number 
less than the entropy rate,
the optimal failure probability goes to $0$ exponentially.
Hence, based on its exponential rate (failure exponent), we can 
roughly estimate the failure probability for a given finite number of copies.
As preceding researches,
Hayashi {\it et al}.\cite{HKMMW}
derived the failure exponent of entanglement concentration
in the i.i.d.\ case based on the method of types, and
Hayashi and Matsumoto \cite{HM} did
that of their universal entanglement concentration protocol.
In this paper, we calculate 
the failure exponent of entanglement concentration
in a more general setting.

In most problems in information theory, 
in the i.i.d.\ case, the correct (or success) probability 
exponentially goes to 0
when the rate is strictly better than the optimal rate.
This exponential rate is called the correct (or success) exponent,
and is one of famouse issues in information theory.
Hayashi {\it et al}.\cite{HKMMW} and
Hayashi and Matsumoto \cite{HM}
treated the success exponent of entanglement concentration in the 
i.i.d.\ case.
This paper proceed to the general sequence of 
partially entangled pure states.

One may think that such an exponential treatment is not essential.
It is, however, more difficult to obtain the 
error and correct exponents asymptotically and tightly
than the asymptotical optimal production rate.
Hence, in order to derive these tight bounds of 
both exponents,
we need better and more simple non-asymptotic evaluations.
That is, such a non-asymptotic evaluation
should be a better and more simple approximation for the optimal value.
Therefore, even though the optimal correct exponent is useless,
the non-asymptotic evaluations used for its derivation
is quite useful.

Furthermore, 
the optimal rates with exponential constraint
are characterized by 
R\'{e}nyi entropy in the i.i.d. case.
In this paper, we derive the same formulas under a
weak assumption for the R\'{e}nyi entropy.
Using these formulas, we characterize 
the optimal rates based on 
the partition function.

Finally, we have to explain our formulation of
entanglement concentration.
There are two formulations in source coding.
One is fixed length, in which
the coding length is fixed, {\it i.e.},
is independent of the input data.
The other is variable-length,
in which the coding length is variable, {\it i.e.},
depends on the input data.
Similarly to source coding,
we can consider two similar formulations in entanglement concentration.
In Bennett {\it et al.}\cite{Ben}'s protocol and Hayashi and 
Matsumoto\cite{HM}'s protocol,
a local measurement is required as the first step,
and the length of the MSE generated finally 
depends on the data of this local measurement.
Hence, their protocol is a variable-length entanglement concentration.

On the other hand,
based on Nielsen's result\cite{N},
Hayashi {\it et al}.\cite{HKMMW} 
discussed entanglement concentration protocols
producing the MES with the fixed size.
Hence, such protocols are called 
fixed-length entanglement concentration,
which are classified into two formulations as follows.
In the first formulation, we produce, without a failure,
an approximately MES from a partially entangled pure state.
Its performance is represented by the size of the MES
and the fidelity between the appropriate MES and the final state.
This kind of entanglement concentration is 
called deterministic fixed-length 
entanglement concentration (DFLEC).
In the other formulation, we produce
an MES itself, allowing a failure probability, from 
a partially entangled pure state.
The performance of this protocol
is evaluated by the size of the MES
and the failure probability.
This protocol is called a probabilistic fixed-length 
entanglement concentration (PFLEC).
Hayashi {\it et al}.\cite{HKMMW} treated these two
formulations in the i.i.d.\ case.
In this paper, we discuss them in a more general model.

This paper is organized as follows.
In section \ref{s2}, 
we give the mathematical definitions 
of the optimal rates with respective conditions,
(constant constraint, exponetial constraint)
for the genereal sequence of partially entangled pure state
in two formulations of FLEC.
As the main results,
characterizations of these quantities are given based on 
information spectrum.
That is, we discover a general relation 
between the performance of entanglement concentration
and the eigenvalues of the reduced density (partially traced state).
In section \ref{s3}, 
the optimal rates of FLECs 
are characterized by the R\'{e}nyi entropy.
In section \ref{s3-2}, 
we apply these formulas to 
the case when the reduced density is given as a thermal state.
In section \ref{s4},
the performances of the two FLEC types in a non-asymptotic case
are characterized by applications of Nielsen's result \cite{N} 
and Lo and Popescu's results\cite{LP}.
In section \ref{s5},
the main result is verified by applying several lemmas described
in section \ref{s4} to an asymptotic case.
In section \ref{s7},
the relation between entanglement concentration and 
random number generation is discussed.
The appendix \ref{as1}
summarizes relations for the quantum analogue of the information 
spectrums based on the original definition\cite{NH},
which are necessary for verifying the main result.

\section{Main results}
\Label{s2}
When the two distinct parties, Alice and Bob, have their respective 
systems ${\cal H}_A$ and ${\cal H}_B$,
the total system is described by the tensor product space
${\cal H}_A\otimes {\cal H}_B$.
In quantum information,
as is mentioned in section \ref{s1},
one of main issues is the characterization of entanglement between these 
distinct parties.
If the state on total syste is a pure state 
$\Phi\in {\cal H}_A\otimes {\cal H}_B$,
it is known that its entanglement between two parties can be characterized 
by the reduced density (partially traced state) $\rho :=
\Tr_B |\Phi\rangle \langle \Phi|$.
In particular, 
if the reduced density $\rho$ is the completely mixed state
$\frac{1}{d_A}I$,
it is called maximally entagled, where $d_A$ denotes the dimension of
the system ${\cal H}_A$.
Hence, if the pure state $\Psi \in {\cal H}_A\otimes {\cal H}_B$
is maximally entangled,
there exist
completely orthogonal basis $\{ e_i\}$ and $\{ e_i'\}$ 
on ${\cal H}_A$ and ${\cal H}_B$, respectively
such that
\begin{align*}
\Psi= \sqrt{\frac{1}{d_A}}
\sum_{i=1}^{d_A}e_i \otimes e_i'.
\end{align*}
While any quantum operation is mathematically described by 
trace-preserving completely positive (TP-CP) map,
in the entanglement concentration
of the initial pure state $\Phi$ 
on distinct two parties ${\cal H}_A$ and ${\cal H}_B$,
our operation is often restricted to
a quantum operation with an LOCC implementation
between ${\cal H}_A$ and ${\cal H}_B$.
Hence, a deterministic fixed-length 
entanglement concentration (DFLEC) is an LOCC quantum operation 
$C$ together with a maximally entangled state $\Psi$,
on a subspace ${\cal H}_A'\otimes {\cal H}_B'$,
{\it i.e.}, it is described as $(C, \Psi)$.
Since this protocol $(C, \Psi)$ transforms
the initial pure state $|\Phi\rangle \langle \Phi|$ 
to the final state $C(\Phi):=
C(| \Phi \rangle \langle \Phi |)$,
its performance is evaluated by
the fidelity
$\langle \Psi | C(\Phi) | \Psi \rangle$ and the size $L(\Psi)$ of $\Psi$,
which equals $H(\Tr_B|\Psi\rangle \langle \Psi|)$.

For a rigid analysis of the probabilistic fixed-length 
entanglement concentration,
we have to discuss a measuring operation
that describes a quantum measurement
with the final state as well as the probability distribution
of the measured data.
The measuring operation 
is given as a CP map valued measure $I=\{ I_i\}_i$ 
whose sum is a TP-CP map; {\it i.e.}, every $I_i$ 
is a CP map, and $\sum_i I_i$ is a TP-CP map.
It is often called an instrument.
When we perform a quantum measurement corresponding to $I=\{ I_i\}_i$ 
on the system with a state $\rho$,
we obtain the measured data $i$ and the final state 
$\frac{1}{\Tr I_i(\rho)}I_i(\rho)$ with the probability $\Tr I_i(\rho)$.
Hence, 
a probabilistic fixed-length entanglement concentration
(PFLEC)
of an initial pure state $\Phi \in {\cal H}_A \otimes {\cal H}_B$
is 
a two-valued instruments $I= \{ I_0, I_1\}$ 
with an LOCC implementation
satisfying that
$I_1(\Phi)
/ \Tr I_1(\Phi)$
is a maximally entangled state $|\Psi\rangle \langle \Psi|$
on a subspace ${\cal H}_A'\otimes {\cal H}_B'$,
where $I_i(| \Phi \rangle \langle \Phi |)$ is abbreviated to $I_i(\Phi)$.
That is, the event $1$ corresponds to success,
and the event $0$ does to failure.
Thus, its performance is characterized by the failure 
probability $\Tr I_0( \Phi)$
and the size $L(I):= L(\Psi)$ of the final maximally entangled state.

Here, we briefly discuss the relation
between two kinds of fixed-length entanglement concentrations.
For any PFLEC $I= \{ I_0, I_1\}$ of $\Phi$,
the pair $(I_1+I_0,I_1( \Phi )
/ \Tr I_1(\Phi))$
becomes a DFLEC and its fidelity between the final state 
and the desired maximally entangled state $I_1(\Phi)
/ \Tr I_1(\Phi)$ is 
greater than 
the success probability of the DFLEC $I= \{ I_0, I_1\}$:
\begin{align}
\Tr \left[(I_1+I_0)(\Phi )
\frac{I_1(\Phi)}
{\Tr I_1(\Phi)}\right]
\ge \Tr I_1(\Phi) . \Label{c51}
\end{align}
That is,
for any a given PFLEC protocol, there exists a DFLEC protocol whose
performance is better than the given PFLEC protocol.

In the quantum system,
if $n$ systems are prepared identically 
to the system ${\cal H}_{A} \otimes {\cal H}_{B}$,
the total system is described by
${\cal H}_A^{\otimes n} \otimes {\cal H}_B^{\otimes n}$.
If the state of every system ${\cal H}_{A} \otimes {\cal H}_{B}$
is the pure state $\Phi$ and if each system is independently prepared,
the state of the total system is 
written by the tensor product pure state $\Phi^{\otimes n}$.
Such a case is called the i.i.d. case.
However, even if the state of each system coincides with each other,
if they are not independent of each other,
the state of total system is not
a tensor product state.
In order to treat such a general case, 
we focus on 
a general sequence of 
the pair of 
the joint system with distinct two parties 
${\cal H}_{A,n} $ and ${\cal H}_{B,n}$ and 
the partially entangled pure state
$\Phi_n \in {\cal H}_{A,n} \otimes {\cal H}_{B,n}$
with an asymptotic situation.
Note that, in this notation, the space ${\cal H}_{A,n}$ and ${\cal H}_{B,n}$
are generalizations
of ${\cal H}_A^{\otimes n}$ and ${\cal H}_B^{\otimes n}$,
and $\Phi_n$ is a generalization
of the $n$-tensor product vector $\Phi^{\otimes n}$.

In order to discuss the asymptotic optimal performance in such a general case,
we optimize
the production rate of MES with three asymptotic 
constraints for the failure probability or fidelity.
Concerning the PFLEC,
we focus on the following conditions:
\begin{itemize}
\item Constant constraint:
The asymptotic failure probability is less than a fixed constant.
\item
Exponential constraint for the failure probability:
When we choose a good DFLEC protocol,
failure probability goes to $0$ exponentially.
Hence, as another criterion, we restrict our DFLEC
satisfying that the exponent of failure probability
is greater than a fixed exponent.
\item
Exponential constraint for the success probability:
If we choose a bad DFLEC,
the success probability goes to $0$ exponentially.
Among such PFLEC protocols,
if this exponent, {\it i.e.}, the success exponent,
is greater, 
the protocol is worse.
Hence, we can consider the optimization of
the production rate of MES with the constraint
that the success exponent is less than 
a fixed exponent.
\end{itemize}
Thus, concerning PFLEC,
we focus on the following values:
\begin{align*}
B_P(\epsilon)&:=
\sup_{\{I^n\}} 
\Bigl\{\varliminf \frac{\log L(I^n) }{n}
\Bigl|
\varlimsup I_0^n( \Phi_n )
\le \epsilon \Bigr\} \\
B_{e,P}(r)&:=
\sup_{\{I^n\}} 
\Bigl\{
\varliminf \frac{\log L(I^n)}{n}
\Bigr|
\varliminf \frac{-1}{n}\log \Tr I_0^n(\Phi_n )
\ge r
\Bigr\} \\
B_{e,P}^*(r)&:=
\sup_{\{I^n\}} \Bigl\{
\varliminf \frac{\log L(I^n)}{n}
\Bigr|
\varlimsup \frac{-1}{n}\log \Tr I_1^n(\Phi_n)
\le r
\Bigr\} .
\end{align*}

In the DFLEC case,
we obtain several criteria
by replacing the success probability in the above discussion
by the fidelity.
That is, we can 
define the following values:
\begin{align*}
B_D(\epsilon)&:=
\sup_{\{(C^n,\Psi_n)\}} 
\Bigl\{\varliminf \frac{1}{n}\log L(\Psi_n) \Bigl|\\
&\hspace{13.6ex} \varliminf \langle \Psi_n | C^n(\Phi_n) | \Psi_n \rangle
\ge 1- \epsilon \Bigr\} \\
B_{e,D}(r)&:=
\sup_{\{(C^n,\Psi_n)\}} 
\Bigl\{
\varliminf \frac{1}{n}\log L(\Psi_n)
\Bigr|\\
&\hspace{5ex} \varliminf \frac{-1}{n}
\log \left(
1- \langle \Psi_n 
| C^n(\Phi_n ) | \Psi_n \rangle
\right)
\ge r
\Bigr\} \\
B_{e,D}^*(r)&:=
\sup_{\{(C^n,\Psi_n)\}} 
\Bigl\{
\varliminf \frac{1}{n}\log L(\Psi_n)
\Bigr| \\
&\hspace{10.5ex}
\varlimsup \frac{-1}{n}\log 
\langle \Psi_n | C^n(\Phi_n) | \Psi_n \rangle
\le r
\Bigr\} .
\end{align*}

Hence, it is trivial from (\ref{c51}) that
\begin{align}
B_1(\epsilon) \ge B_2(\epsilon) , \quad
B_{e,D}(r) \ge B_{e,2}(r) , \quad
B_{e,D}^*(r) \ge B_{e,2}^*(r) .
\end{align}
In this paper, we treat a quantum analogue of information spectrums
to analyze the above values.
For such an analysis, we need the following definitions.
For a self-adjoint operator $X$,
we can denote the projection
$\sum_{x_i \ge c } E_i$ by 
$\{ X \ge c \}$,
where the spectral decomposition is given by
$X= \sum_i x_i E_i$.
We can define the projections
$\{ X \,> c \},\{ X \,< C \}$, $\{ X \le c \}$, {\it etc}. in a similar
manner.
Let $\rho_n$ be the reduced density
$\Tr _{{\cal H}_{B,n}}| \phi_n \rangle \langle \phi_n |$
and define 
\begin{align*}
K(a)&:=
\varlimsup \Tr \rho_n \{ \rho_n -e^{-na}\ge 0\} \\
\underline{\zeta}^c(a)&:=
\varliminf \frac{-1}{n}\log \Tr \rho_n\{ \rho_n - e^{-na} \,> 0 \} .
\end{align*}
When the limit 
\begin{align}
\lim \frac{-1}{n}\log \Tr \rho_n\{ \rho_n - e^{-na} \,< 0 \} \Label{65}
\end{align}
exists, we denote it by $\zeta(a)$.
These definitions can also be written as
\begin{align}
K(a)&=
\varlimsup p_n 
\left\{ \frac{-1}{n}\log p_{n,i} \le a \right\} \Label{691}\\
\underline{\zeta}^c(a) &= 
\varliminf \frac{-1}{n}\log p_n 
\left\{ \frac{-1}{n}\log p_{n,i} \le a \right\} \Label{69}\\
\zeta(a) &=
\lim \frac{-1}{n}\log p_n 
\left\{ \frac{-1}{n}\log p_{n,i} \,> a \right\} ,\Label{70}
\end{align}
where every
$p_{n,i}$ is an eigenvalue of $\rho_n$
and can be regarded as a probability distribution.
Hence, the quantity $K(a),\underline{\zeta}^c(a)$, and
$\zeta(a)$ denotes 
the degree of concentration 
of the $e^{na}$-dimensional subspace.
Note that the function $\underline{\zeta}^c(a)$
decreases monotonically, while 
the function $\zeta(a)$ increases monotonically.
Indeed, in the classical case, the value $K(a)$
gives the asymptotic performances of 
fixed-length source coding\cite{Han_source}
and uniform random number generation\cite{V-V,Han_book}
with asymptotic constant constraint.
Moreover, the quantities $\underline{\zeta}^c(a)$ and 
$\zeta(a)$ gives the 
asymptotic optimal performance of source coding 
with the exponential constraint\cite{Han_source}
and that of simulation of random process with 
KL divergence criterion\cite{SV}.
As is mentioned in section \ref{s7}, 
$\zeta(a)$ gives the asymptotic optimal performance 
of intrinsic randomness 
with KL divergence criterion\cite{H-second}.

As is mentioned in the following main theorem,
the optimal production rate of MES 
can be characterized by how densely
the eigen values of 
the reduced density matrix
concentrate a small space.
\begin{thm}\Label{t1}
Without any assumption, for every $\epsilon \in [0,1]$ we have 
\begin{align*}
B_{D}(\epsilon)&=
B_{P}(\epsilon)=\sup_R\{ R | K(R) \le \epsilon\} \\
B_{e,D}(r) &= B_{e,P}(r) =
\sup_R \{ R | \underline{\zeta}^c (R) \ge r \}.
\end{align*}
When the limit (\ref{65}) exists
and there exists a real number $a$ such that 
$\zeta(a)\le \underline{\zeta}^c(a)$,
we have
\begin{align*}
B_{e,D}^*(r)&=
\sup_a \left\{ a -r \left| 
\inf_{a'} \left\{\zeta(a')- \frac{a'}{2}\right| 
a' \le a \right\}+ \frac{a}{2}
\le r \right\} \\
&=
\sup_a \left\{  \frac{a}{2} -
\inf_{a'} \left\{\left.\zeta(a')- \frac{a'}{2}\right| 
a' \le a \right\}
 \right| \\
&\hspace{15ex}\left.\inf_{a'} \left\{\left.\zeta(a')- \frac{a'}{2}\right| 
a' \le a \right\}+ \frac{a}{2}
\le r \right\}  
\\
B_{e,P}^*(r)&=
\sup_a\{ a- \zeta (a) | \zeta(a) \le r\}.
\end{align*}
\end{thm}
This theorem is proved in section \ref{s5}
after preparing the appropriate discussion.
\begin{rem}
As is mentioned in Nagaoka and Hayashi\cite{NH}
the quantum versions of $K(a),\underline{\zeta}^c(a)$, and
$\zeta(a)$ give the asymptotic performances of 
fixed-length source coding.
In particular, 
the optimal rate with the constraint for the constant error exponent
is given as
\begin{align}
\sup_a\{ a- \zeta (a) | \zeta(a) < r\},
\end{align}
which is almost similar to $B_{e,P}^*(r)$.
For a proof only of the classical case,
see Han \cite{Han_source}.
For a proof in the classical and quantum case,
see Nagaoka and Hayashi \cite{NH}.
\end{rem}
\section{Asymptotic formulas based on R\'{e}nyi entropy}\Label{s3}
In the classical and quantum fixed-length source coding of 
i.i.d. information source,
it is known that
the optimal rate with the 
constant constraint for error exponent
is described by the R\'{e}nyi entropy
$\psi(s):= \log \sum_i p_i^s$\cite{CK}.
Concerning FLEC of the i.i.d. source, 
as is described in Theorem \ref{tH},
Hayashi {\it et al.}\cite{HKMMW} showed 
that this kinds of optimal rates 
can be described by the R\'{e}nyi entropy.
In this section, using Theorem \ref{t1},
we derive 
the same formula in a more general setting.
\begin{thm}\Label{tH}{\bf 
 Hayashi {\it et al.}\cite{HKMMW}}
When $\rho_n = \rho^{\otimes n}$, the 
relations
\begin{align}
B_{D}(\epsilon)&=
B_{P}(\epsilon)=H(\rho) , \quad 
\forall \epsilon \hbox{ such that }
1 \,> 
\epsilon \ge 0\Label{62}
\\
B_{e,D}(r)&=
B_{e,P}(r)=
\sup_{s \ge 1 } \frac{r+\psi(s)}{1-s}
\Label{60}\\
B_{e,P}^*(r)&=
\min_{0 \le s \le 1}
\frac{sr+\psi(s)}{1-s}
\Label{61} \\
B_{e,D}^*(r)&=
\left\{
\begin{array}{ll}
\displaystyle \min_{0 \le s \le 1}
\frac{sr+\psi(s)}{1-s}
& \hbox{ if } r \le
-\frac{1}{2}\psi'\left(\frac{1}{2}\right)
-\psi\left(\frac{1}{2}\right) \\
2\psi\left(\frac{1}{2}\right)+ r
& \hbox{ otherwise}
\end{array}
\right. \Label{77}
\end{align}
hold, where
\begin{align*}
H(\rho):= - \Tr \rho \log \rho ,\quad
\psi(s):= \log \Tr \rho^s .
\end{align*}
In particular, the above formulas of 
some special cases are written as
\begin{align*}
B_{e,D}(r)=B_{e,P}(r)=H_{\infty} & \hbox{ if }
r \ge H_{\infty}=\lim _{s \to \infty}-\psi'(s) \\
B_{e,P}^*(r)= \psi(0) &\hbox{ if }
r \ge -\psi'(0)-\psi(0),
\end{align*}
where
\begin{align*}
H_{\infty} :=
\lim_{s \to \infty}\frac{-\psi(s)}{s}.
\end{align*}
\end{thm}
The following is 
the generalization of 
the above theorem.
\begin{thm}\Label{t-g}
Letting $\psi_n(s):= \log \Tr \rho_n^s$,
we assume that 
the limit $\overline{\psi}(s):= 
\lim_n \frac{\psi_n(s)}{n}$ exists 
and that its first and second derivatives
$\overline{\psi}'(s)$ and $\overline{\psi}''(s)$ exist
for $s \in (0,1) \cup (1,\infty)$.
Then, 
\begin{align}
\overline{H}_- &\le 
B_{D}(\epsilon)
=
B_{P}(\epsilon)\le\overline{H}_+\Label{62-g}\\
B_{e,D}(r)&=
B_{e,P}(r)=
\sup_{s \ge 1 } \frac{r+\overline{\psi}(s)}{1-s}
\Label{60-g}\\
B_{e,P}^*(r)&=
\min_{0 \le s \le 1}
\frac{sr+\overline{\psi}(s)}{1-s}
\Label{61-g} \\
B_{e,D}^*(r)&=
\left\{
\begin{array}{ll}
\displaystyle \min_{0 \le s \le 1}
\frac{sr+\overline{\psi}(s)}{1-s}
& \hbox{ if } r \le
-\frac{1}{2}\overline{\psi}'\left(\frac{1}{2}\right)
-\overline{\psi}\left(\frac{1}{2}\right) \\
2\overline{\psi}\left(\frac{1}{2}\right)+ r
& \hbox{ otherwise},
\end{array}
\right. \Label{77-g}
\end{align}
where
\begin{align*}
\overline{H}_-:=
-\overline{\psi}'(1+0), \quad
\overline{H}_+:=
-\overline{\psi}'(1-0).
\end{align*}
In particular, we have
\begin{align*}
B_{e,D}(r)=B_{e,P}(r)=\overline{H}_{\infty} & \hbox{ if }
r \ge \overline{H}_{\infty}=\lim _{s \to \infty}-\overline{\psi}'(s) \\
B_{e,P}^*(r)= \overline{\psi}(0) &\hbox{ if }
r \ge -\overline{\psi}'(+0)-\overline{\psi}(0),
\end{align*}
where $\overline{H}_{\infty} :=
\lim_{s \to \infty}\frac{-\overline{\psi}(s)}{s}$.
\end{thm}
The equations 
(\ref{60}), (\ref{61}), and (\ref{77})
follow from 
the equations (\ref{60-g}), (\ref{61-g}), and (\ref{77-g}).
The equation (\ref{62}) follows from
the equation (\ref{62-g}).
Hence, Theorem \ref{t-g} can be regarded as
a generalization of Theorem \ref{tH}.
Since Hayashi {\it et al.} \cite{HKMMW}used the method of type,
they proved Theorem \ref{tH} only in the finite-dimensional case.
Hence, its infinite-dimensional case is proved by this paper first time.

\begin{rem}
Under the same assumption as Theorem \ref{t-g},
we can similarly prove that
\begin{align*}
\sup_a\{ a- \zeta (a) | \zeta(a) < r\}
=
\min_{0 \le s \le 1}
\frac{sr+\overline{\psi}(s)}{1-s},
\end{align*}
which gives the 
optimal rate with the constant constraint for error exponent
in the fixed-length source coding.
\end{rem}

\begin{proof}
As is discussed in Appendix \ref{GE},
G\"{a}rtner-Ellis theorem \cite{DZ} yields that
\begin{align}
\zeta(a)
&=
\left\{
\begin{array}{cl}
0 & \hbox{ if }a \le \overline{H}_+  \\
\displaystyle  \sup_{0 \le s \le 1}
(1-s)a -\overline{\psi}(s)>0  &\hbox{ if }
\overline{H}_+ < a < - \overline{\psi}'(0) 
\end{array}
\right.
\Label{66-a}\\
\underline{\zeta}^c(a)
&=
\left\{
\begin{array}{cl}
0 & \hbox{ if } \overline{H}_- \le a \\
\displaystyle  \sup_{s \ge 1}
(1-s) a - \overline{\psi}(s)>0 &\hbox{ if }
\overline{H}_{\infty}< a < \overline{H}_-  \\
\infty & \hbox{ if }
a \,< \overline{H}_{\infty}.
\end{array}
\right. \Label{66}
\end{align}
Note that 
\begin{align}
\zeta(- \overline{\psi}'(+0)-0) &=
- \overline{\psi}'(+0) -\overline{\psi}(0) \Label{3-22}\\
\underline{\zeta}^c(\overline{H}_\infty+0)&=
\overline{H}_\infty .
\end{align}
Moreover, it follows from the discussion in Appendix \ref{GE}
that $\overline{\psi}(s)$ is convex.
Since $\overline{\psi}''(s)$ exists 
for $s \in (0,1) \cup (1,\infty)$,
we have \begin{align}
\overline{\psi}''(s)\ge 0 
\quad s \in (0,1) \cup (1,\infty).
\Label{ato}
\end{align}

For any real number $a$ satisfying
$ \overline{H}_{\infty} \le a \le - \overline{\psi}'(+0)$, 
we define 
$s(a)$ by
\begin{align}
a &= -\overline{\psi}'(s(a)). \Label{73} 
\end{align}
Hence, equations (\ref{66-a}) and (\ref{66}) yield that
\begin{align}
\zeta(a)
&=
\left\{
\begin{array}{cl}
0 & \hbox{ if }a \le \overline{H}_+  \\
(1-s(a))a - \overline{\psi}(s(a))
&\hbox{ if }
\overline{H}_+ < a < - \overline{\psi}'(0) 
\end{array}
\right.
\Label{66-a-1}\\
\underline{\zeta}^c(a)
&=
\left\{
\begin{array}{cl}
0 & \hbox{ if } \overline{H}_- \le a \\
(1-s(a))a - \overline{\psi}(s(a))
&\hbox{ if }
\overline{H}_{\infty}< a < \overline{H}_-  \\
\infty & \hbox{ if }
a \,< \overline{H}_{\infty}.
\end{array}
\right. \Label{66-1}
\end{align}

First, we prove (\ref{60-g}) for the case in which
$r \,< \overline{H}_{\infty}$.
In this case, we can define $a_r$ and $s_r$ by
$\underline{\zeta}^c(a_r)=r$ and $s_r:=s(a_r)$.
Thus, we have
\begin{align}
(1-s_r)a_r - \overline{\psi}(s_r)&= r \Label{b1}\\
-(1-s_r)\overline{\psi}'(s_r) - \overline{\psi}(s_r)&= r . \Label{b2}
\end{align}
Using (\ref{b1}),
we can calculate $B_{e,D}(r)$ and $B_{e,P}(r)$ as
\begin{align*}
B_{e,D}(r)=B_{e,P}(r)=a_r=
\frac{r+\overline{\psi}(s_r)}{1-s_r}.
\end{align*}
The derivative of the function
$f_1(s):= \frac{r+\overline{\psi}(s)}{1-s}(s \ge 1)$
is given by
\begin{align*}
f_1'(s)=\frac{\overline{\psi}'(s)(1-s)+r +\overline{\psi}(s)}{(1-s)^2}.
\end{align*}
From (\ref{b2}), the equation $f_1(s_r)'=0$ holds.
The derivative of the numerator of $f_1'(s)$ is 
\begin{align*}
\left(\overline{\psi}'(s)(1-s)+r +\overline{\psi}(s)\right)'=
\overline{\psi}''(s)(1-s) \le 0 ,
\end{align*}
the final inequality inequality follows from (\ref{ato}).
Therefore, 
$B_{e,D}(r)=B_{e,P}(r)=f_1(s_r)= \max_{s \ge 1} f_1(s)$.

Next, we prove (\ref{60-g}) for the case in which
$r \ge \overline{H}_{\infty}$.
From (\ref{66}), if $a \,> \overline{H}_\infty$, 
then $\underline{\zeta}^c(a) \,< r$.
Otherwise, $\underline{\zeta}^c(a) \ge r$.
Thus, $B_{e,D}(r)=B_{e,P}(r)=\overline{H}_\infty$.
Since the numerator of $f_1'(s)$ equals
\begin{align*}
r+\overline{\psi}'(s)(1-s) +\overline{\psi}(s) 
= r- \underline{\zeta}^c(-\overline{\psi}'(s)) \,> 0,
\end{align*}
we obtain $f_1'(s) \,> 0$.
Therefore,
\begin{align*}
\sup_{s \ge 1}
\frac{r+\overline{\psi}(s)}{1-s} =
\lim_{s \to \infty} \frac{r+\overline{\psi}(s)}{1-s}
= \overline{H}_\infty.
\end{align*}

Proceeding to (\ref{61-g}) for the case in which
$r \,< -\overline{\psi}'(+0)-\overline{\psi}(0)$,
we define $a_r$ and $s_r$ by
$\zeta(a_r)=r$ and $s_r:=s(a_r)$.
Thus, we have
\begin{align}
(1-s_r)a_r - \overline{\psi}(s_r)&= r \Label{b3}\\
-(1-s_r)\overline{\psi}'(s_r) - \overline{\psi}(s_r)&= r . \Label{b4}
\end{align}
Using (\ref{b3}),
we can calculate $B_{e,P}^*(r)$:
\begin{align*}
B_{e,P}^*(r)=a_r -r =
\frac{s_r r+\overline{\psi}(s_r)}{1-s_r}.
\end{align*}
The derivative of the function
$f_2(s):= \frac{s r+\overline{\psi}(s)}{1-s}(0\,< s \,< 1)$
is given by
\begin{align*}
f_2'(s)=\frac{\overline{\psi}'(s)(1-s)+r +\overline{\psi}(s)}{(1-s)^2}.
\end{align*}
From (\ref{b4}), the equation $f_2(s_r)'=0$ holds.
The derivative of the numerator of $f_2'(s)$ is given by
\begin{align*}
\left(\overline{\psi}'(s)(1-s)+r +\overline{\psi}(s)\right)'=
\overline{\psi}''(s)(1-s) \ge 0 
\end{align*}
because of (\ref{ato}).
Therefore, 
$B_{e,P}^*(r)=f_2(s_r)= \min_{s \ge 1} f_2(s)$.

Next, we prove (\ref{61-g}) for the case in which
$r \ge -\overline{\psi}'(+0)-\overline{\psi}(0)$.
If $a \,< -\overline{\psi}'(+0)$, then $\zeta(a) 
\,< r$.
Otherwise, $\zeta(a) \,> r$.
Thus, it follows from (\ref{3-22}) that $B_{e,P}^*(r)=\lim_{\epsilon \to +0}
(- \overline{\psi}'(+0)- \epsilon)
-
\zeta(- \overline{\psi}'(+0)- \epsilon)
= -\overline{\psi}'(+0) - (-\overline{\psi}'(+0)- \overline{\psi}(0))
=\overline{\psi}(0)$.
Since the numerator of $f_2'(s)$ is
\begin{align*}
r+\overline{\psi}'(s)(1-s) +\overline{\psi}(s) 
= r- \underline{\zeta}^c(-\overline{\psi}'(s)) \,> 0,
\end{align*}
then $f_2'(s) \,> 0$.
Therefore,
\begin{align*}
\min_{0 \le s \le 1}
\frac{s r+\overline{\psi}(s)}{1-s} =
\lim_{s \to 0} \frac{s r+\overline{\psi}(s)}{1-s}
= \overline{\psi}(0).
\end{align*}

Next, we prove (\ref{77-g}).
We can calculate the derivative of
$\zeta(a)- \frac{a}{2}$ as
\begin{align*}
& \left(
\zeta(a)- \frac{a}{2}
\right)'
=
1- s(a) -s'(a)a - \overline{\psi}'(s(a))s'(a)
- \frac{1}{2} \\
=&
1- s(a) -s'(a)a  + s'(a)a  - \frac{1}{2}= 
\frac{1}{2} - s(a).
\end{align*}
This derivative is $0$ if and only if
$s(a)= \frac{1}{2}$, {\it i.e.},
$a= - \overline{\psi}' \left(\frac{1}{2}\right)$.
The second derivative is calculated as
\begin{align}
\left(
\zeta(a)- \frac{a}{2}
\right)''
=- s'(a)
= \frac{1}{\overline{\psi}''(s(a))}>0, \Label{78}
\end{align}
where the final equation follows from 
$1= \overline{\psi}''(s(a)) s'(a)$ which can be
derived from (\ref{73}).
Thus, the function $a \mapsto
\zeta(a)- \frac{a}{2}$ is strictly convex, and
its minimum value 
equals $\zeta\left(
- \overline{\psi}' \left(\frac{1}{2}\right)
\right) + \frac{1}{2} \overline{\psi}' \left(\frac{1}{2}\right)$,
which is attained 
at $a= - \overline{\psi}' \left(\frac{1}{2}\right)$.
Hence, we have
\begin{align*}
\inf_{a'}\left\{ \left. \zeta(a') - \frac{a'}{2}
\right|
a' \le a \right\}=
\left\{
\begin{array}{ll}
\zeta(a) -\frac{a}{2} & \hbox{ if }
a \le - \overline{\psi}' \left(\frac{1}{2}\right)
\\
 -\overline{\psi}\left(\frac{1}{2}\right) &\hbox{ if }
a > - \overline{\psi}' \left(\frac{1}{2}\right),
\end{array}
\right.
\end{align*}
where we use 
the equation 
$\zeta\left(- \overline{\psi}' \left(\frac{1}{2}\right)\right) 
= -\overline{\psi}\left(\frac{1}{2}\right)
- \frac{1}{2}\overline{\psi}' \left(\frac{1}{2}\right) $,
which follows from (\ref{66-a-1}).
Since
$\zeta(- \overline{\psi}' (1/2))= 
-\overline{\psi}\left(\frac{1}{2}\right) 
-\frac{1}{2}\overline{\psi}' \left(\frac{1}{2}\right)$,
we have
\begin{align*}
&\sup_{a \le - \overline{\psi}' (1/2)}
\left\{ a- r| \zeta(a)\le r \right\}\\
=&
\left\{
\begin{array}{ll}
a_r - r
& \hbox{ if } r \le -\overline{\psi}\left(\frac{1}{2}\right) 
-\frac{1}{2}\overline{\psi}' \left(\frac{1}{2}\right)\\
- \overline{\psi}' (1/2) -r
& \hbox{ if } r > -\overline{\psi}\left(\frac{1}{2}\right) 
-\frac{1}{2}\overline{\psi}' \left(\frac{1}{2}\right).
\end{array}
\right.
\end{align*}
Remember that 
$a_r$ is defined such that $\zeta(a_r)=r$.
Moreover, we have
\begin{align*}
&\sup_{a > - \overline{\psi}' (1/2)}
\left\{ a- r\left| -\overline{\psi}\left(\frac{1}{2}\right) +\frac{a}{2}
\le r \right.\right\}\\
=&
\left\{
\begin{array}{ll}
0
& \hbox{ if } r \le -\overline{\psi}\left(\frac{1}{2}\right) 
-\frac{1}{2}\overline{\psi}' \left(\frac{1}{2}\right)\\
2 \overline{\psi} \left(\frac{1}{2}\right) +r
& \hbox{ if } r > -\overline{\psi}\left(\frac{1}{2}\right) 
-\frac{1}{2}\overline{\psi}' \left(\frac{1}{2}\right).
\end{array}
\right.
\end{align*}
Therefore,
\begin{align*}
B_{e,D}^*(r)=&
\sup_a 
\left\{ a- r \left|
\inf_{a'} \left\{\left.\zeta(a')- \frac{a'}{2}\right| 
a' \le a \right\}+ \frac{a}{2}
\le r 
\right.\right\}\\
= &
\max\left\{
\sup_{a\le - \overline{\psi}'(1/2)} 
\left\{ a- r \left|
\zeta(a)\le r \right.\right\}, \right.\\
& \hspace{6ex} \left. \sup_{a> - \overline{\psi}'(1/2)} 
\left\{ a- r \left|
 -\overline{\psi}\left(\frac{1}{2}\right) +\frac{a}{2}
\le r \right.\right\}
\right\}\\
=&
\left\{
\begin{array}{ll}
a_r -r
& \hbox{ if } r \le -\overline{\psi}\left(\frac{1}{2}\right) 
-\frac{1}{2}\overline{\psi}' \left(\frac{1}{2}\right)\\
2 \overline{\psi} \left(\frac{1}{2}\right) +r
& \hbox{ if } r > -\overline{\psi}\left(\frac{1}{2}\right) 
-\frac{1}{2}\overline{\psi}' \left(\frac{1}{2}\right).
\end{array}
\right.
\end{align*}
Using a discussin similar to (\ref{61-g}),
we can show (\ref{77-g}).
\end{proof}
\section{Correlated system}\Label{s3-2}
In this section,
we consider the application of Theorem \ref{t-g} to 
correlated systems.
As an example,
the initial state is assumed to be a ground state with the Hamiltonian
$\sum_i H_{i}+ H_{i,i+1}$
on the system $({\cal H}_A \otimes {\cal H}_B)^{\otimes n}$,
where 
$H_i$ is the Hamiltonian of the $i$-th joint system between 
$A$ and $B$,
and $H_{i,i+1}$ is its interaction term between the $i$-th and 
$i+1$-th systems.
However, it is not so easy to 
calculate $\overline{\psi}(s)$ in this case.
Hence, we focus on a more ideal case.

Assume that the total system 
$({\cal H}_A \otimes {\cal H}_B)^{\otimes n}$ is isolated from other systems.
We also assume that
the system ${\cal H}_B^{\otimes n}$ is sufficiently large,
and the interaction between 
the system ${\cal H}_A^{\otimes n}$
and the system ${\cal H}_B^{\otimes n}$
is ideal so that
the system ${\cal H}_B^{\otimes n}$
can be regarded as the heat bath of the system ${\cal H}_A^{\otimes n}$.
Now, we suppose that
the Hamiltonian $\sum_i H_{A,i}+ H_{A,i,i+1}$
on the system ${\cal H}_A^{\otimes n}$.
Hence, 
the state of the total system is pure,
and
the reduced density on $A$ is the thermal state with the
Hamiltonian $\sum_i H_{A,i}+ H_{A,i,i+1}$.
Now, we define the partition function as
\begin{align}
\Xi(\beta):= \lim \frac{1}{n}
\log \Tr \exp (\beta \sum_i H_{A,i}+ H_{A,i,i+1}).
\end{align}
Thus, when the inverse temperature
is $\beta_0$ and the partition function is continuous and 
differentiable,
the $\overline{\psi}(s)$ can be calculated as
\begin{align*}
&\overline{\psi}(s)= \lim \frac{1}{n}
\log \Tr 
\left(
\frac{\exp (\beta_0 \sum_i H_{A,i}+ H_{A,i,i+1})}
{\Tr \exp (\beta_0 \sum_i H_{A,i}+ H_{A,i,i+1})}\right)^s\\
=&
\lim \frac{1}{n}
\log \Tr \exp (s \beta_0 \sum_i H_{A,i}+ H_{A,i,i+1})
- s \Xi(\beta_0)\\
=&\Xi(s \beta_0)- s \Xi(\beta_0).
\end{align*}
Hence, 
\begin{align*}
B_{D}(\epsilon)&=B_{P}(\epsilon)= 
-\beta_0 \Xi'( \beta_0)+ \Xi(\beta_0)\\
B_{e,D}(r)&=
B_{e,P}(r)=
\sup_{s \ge 1 } \frac{r+\Xi(s \beta_0)- s \Xi(\beta_0)}{1-s}
\\
B_{e,P}^*(r)&=
\min_{0 \le s \le 1}
\frac{sr+\Xi(s \beta_0)- s \Xi(\beta_0)}{1-s}
\\
B_{e,D}^*(r)&=
\left\{
\begin{array}{ll}
\displaystyle \min_{0 \le s \le 1}
\frac{sr+\Xi(s \beta_0)- s \Xi(\beta_0)}{1-s}
& \hbox{ if } r \le
r_{1/2}\\
2\Xi\left( \frac{\beta_0}{2}\right)
-\Xi\left( \beta_0\right)+ r
& \hbox{ otherwise},
\end{array}
\right. 
\end{align*}
where
\begin{align*}
r_{1/2}:=
-\frac{\beta_0}{2}
\Xi'\left( \frac{\beta_0}{2}\right)+ \Xi(\beta_0)- \Xi\left( \frac{\beta_0}{2}\right).
\end{align*}
Note that the above formulas are based only on the
partition function.
Hence, it is expected to apply them to other cases.
Moreover, we can derive similar formulas
concerning classical and quantum fixed-length source coding.
\section{Non-asymptotic theory}\Label{s4}
In order to derive general asymptotic formulas based on 
the quantum information spectrums,
we need to prepare approximate formulas
for non-asymptotic setting based on
the form of the reduced density $\rho$.
For this purpose, we focus on 
majorization, because it gives
a necessary 
and sufficient condition for the possibility
of transforming from 
a partially entangled pure state $|\Phi_1 \rangle \langle \Phi_1|$
to another entangled pure state $|\Phi_2 \rangle \langle \Phi_2|$ 
by using LOCC between the two parties
${\cal H}_A$ and ${\cal H}_B$\cite{N}.
Suppose that $p=(p_1, \ldots, p_d)$ and 
$q=(q_1, \ldots, q_d)$ are probability distributions.
The probability $p$ majorizes $q$,
(equivalently $q$ is majorized by $p$),
written $p \succeq q$,
if for each $k$ in the range 
\begin{align*}
\sum_{j=1}^k p^{\downarrow}_j
\ge 
\sum_{j=1}^k q^{\downarrow}_j.
\end{align*}
The elements indicated by $\downarrow$
are taken in descending order;
for example, 
$p^{\downarrow}_1$ is the largest element
in $(p_1, \ldots, p_d)$.
The majorization relation
is a partial order.
To discuss entanglement transformation,
we need to treat probability distributions
consisting of eigenvalues of a reduced density $\rho$.
The reduced density $\rho$ majorizes another
reduced density $\sigma$
written $\rho \succeq \sigma$, if
the probability distribution $p(\rho)$
consisting of eigenvalues of a reduced density $\rho$
majorizes the probability distribution $p(\sigma)$ defined by the other
reduced density $\sigma$.
In particular, the reduced density $\rho$ strongly majorizes another
reduced density $\sigma$,
written $\rho \succ \sigma$,
if $p(\rho) \succeq p(\sigma)$
and if the eigenvector 
corresponding to $p(\rho)^{\downarrow}_j$
coincides with the eigenvector 
corresponding to $p(\sigma)^{\downarrow}_j$.
That is, this condition requires 
that there exists a common basis diagonalizing 
$\rho$ and $\sigma$.
For more information about majorization, 
please see Bhatia's text book\cite{B}.
Using these notations, we can describe Nielsen's 
condition for 
LOCC transformation as follows.
\begin{lem}\Label{l01}{\rm\bf Nielsen\cite{N}}
We can transform an entangled state $\Phi$ to 
another entangled state $\Psi$ 
by LOCC
if and only if
$\sigma \succeq \rho$,
where
$\rho$ ($\sigma$) is the 
reduced density (partially traced state) of 
$\Phi$ ($\Psi$), respectively.
\end{lem}

Therefore, by using the above Nielsen's Lemma,
the optimal performance of DFLEC,
{\it i.e.}, the maximum fidelity
can be evaluated based on majorization as follows.
\begin{lem}\Label{l1}
Let $\sigma$ be the reduced density of 
a given pure state $\Psi $, and $\rho$ be the reduced density of 
the given initial pure state $\Phi$.
Then, we have 
\begin{align}
\max_C \langle \Psi | C( \Phi )|
\Psi \rangle=
\max_{\rho' \succeq \rho}
\max_{U: \hbox{\rm unitary}}
\left(\Tr \sqrt{\rho'} \sqrt{\sigma}U \right)^2 \Label{le01},
\end{align}
where 
the quantum operation $C$ runs over all quantum operations with LOCC 
in the maximum of LHS.
If $\Psi$ is a maximally entangled state with the size $L$,
{\it i.e.}, the operator $T:= L \sigma$ is 
a projection with the rank $L$,
then the relation
\begin{align}
\max_{\rho' \succeq \rho}
\max_{U: \hbox{\rm unitary}}
\left(\Tr \sqrt{\rho'} \sqrt{\sigma}U \right)^2
=
\max_{\rho' \succeq \rho}
\frac{\left(\Tr \sqrt{\rho'} T\right)^2}{L} \Label{le1}
\end{align}
holds.
\end{lem}
\begin{proof}
For any pure state $\Psi,\Phi$,
we have 
\begin{align*}
\langle \Psi| \Phi \rangle=
\Tr_{{\cal H}_A}\sqrt{\rho}\sqrt{\sigma}U_2^* U_1 ,
\end{align*}
where  two unitaries $U_1$ and $U_2$ are defined as 
\begin{align*}
U_1 \rho U_1^* & = 
\Tr_{{\cal H}_A}| \Phi \rangle \langle \Phi| , \quad
U_2 \sigma U_2^*= 
\Tr_{{\cal H}_A}| \Psi \rangle \langle \Psi| .
\end{align*}
Using Lemma \ref{l01},
we can prove (\ref{le01}).
Next, we choose normalized basis 
$\{e_i\}_{i=1}^L$ and $\{f_i\}_{i=1}^L$ as
\begin{align*}
T= \sum_{i=1}^L | e_i \rangle \langle e_i |, \quad
f_i := U^* e_i.
\end{align*}
Using Schwartz inequality twice,
we have 
\begin{align*}
& \Tr \sqrt{\rho'}T U
= \sum_{i=1}^L \langle f_i | \sqrt{\rho'}| e_i \rangle\\
\le & \sum_{i=1}^L \sqrt{\langle f_i | \sqrt{\rho'}| f_i \rangle}
\sqrt{\langle e_i | \sqrt{\rho'}| e_i \rangle} \\
\le & \sqrt{ \sum_{i=1}^L \langle f_i | \sqrt{\rho'}| f_i \rangle}
\sqrt{ \sum_{i=1}^L \langle e_i | \sqrt{\rho'}| e_i \rangle}.
\end{align*}
Since 
\begin{align*}
\sum_{i=1}^L \langle f_i | \sqrt{\rho'}| f_i \rangle ,
\sum_{i=1}^L \langle e_i | \sqrt{\rho'}| e_i \rangle
\le \max_{V:\hbox{unitary}} \Tr V \sqrt{\rho'} V^* T,
\end{align*}
we obtain
\begin{align*}
\max_{U, V:\hbox{unitary}} \Tr V \sqrt{\rho'} V^* T U =
\max_{V:\hbox{unitary}} \Tr V \sqrt{\rho'} V^* T.
\end{align*}
Therefore, the equation
\begin{align}
\max_{\rho' \succeq \rho} \max_{U:\hbox{unitary}} \Tr \sqrt{\rho'}T U
= \max_{\rho' \succeq \rho}\Tr \sqrt{\rho'} T \Label{ll1}
\end{align}
holds because $U\rho U^* \succeq \rho$.
Equations (\ref{le01}) and (\ref{ll1})
guarantee (\ref{le1}).
\end{proof}
However, it is not easy to directly connect the above lemma to 
the information spectrum.
Hence, we prepare the following lemma for the
evaluation of the RHS of (\ref{le1}).
This lemma plays an important role in the converse part of 
the main theorem.
\begin{lem}\Label{l3}
When a projection $T$ and an integer $M$ satisfy 
$\Tr T \ge M$,
and the two reduced densities $\rho'$ and $\rho$ satisfy
$\rho' \succeq \rho$, the inequality
\begin{align}
& \Tr \sqrt{\rho'} T  \nonumber \\ 
\le &
\sqrt{\Tr \left\{ \rho \ge \frac{1}{M}\right\}}
\sqrt{\Tr \rho \left\{ \rho \ge \frac{1}{M}\right\}}\nonumber\\
&+ \sqrt{\Tr T - \Tr \left\{ \rho \ge \frac{1}{M}\right\}}
\sqrt{\Tr \rho \left\{ \rho \,< \frac{1}{M}
\right\}}\Label{le3}
\end{align}
holds.
\end{lem}
\begin{proof}
Assume that $\Tr T = N (\ge M)$.
Without loss of generality,
we can assume that $\rho' \succeq \rho$.
Let us diagonalize $\rho$ and $\rho'$ as 
$\rho= \sum_i s_i |e_i \rangle \langle e_i |$ and
$\rho'= \sum_i s_i' |f_i \rangle \langle f_i |$,
where $s_i \ge s_{i+1}, s_i' \ge s_{i+1}'$.
The inequality
$\Tr \sqrt{\rho'} T \le
\sum_{i=1}^N \sqrt{s_i'}$ holds.
We define the probability distribution $\{ s_{i,N} \}$ 
and $i_N$ as
\begin{align*}
&\{ s_{i,N} \}:=
\arg \max _{\{ s_{i}' \}}
\left\{\left. \sum_{i=1}^N \sqrt{s_i'}\right|
 \{s_i' \} \succeq \{s_i\}\right\},\\
&s_{i_N} \ge \frac{1}{N} \,> s_{i_N+1}.
\end{align*}
Similarly to $i_N$, we can define $i_M$.
Since the function $x \mapsto \sqrt{x}$ is
concave, we can prove that $s_i = s_{i,N} $
for $i \le i_N$.
Since $i_M \le i_N$,
\begin{align*}
&\sum_{i=1}^{i_M}\sqrt{s_{i,N}}
=\sum_{i=1}^{i_M}\sqrt{s_{i}}\\
\le &\sqrt{i_M}\sqrt{\sum_{i=1}^{i_N}s_i}
=\sqrt{\Tr \left\{ \rho \ge \frac{1}{M}\right\}}
\sqrt{\Tr \rho \left\{ \rho \ge \frac{1}{M}\right\}},\\
&\sum_{i=i_M+1}^{N}\sqrt{s_{i,N}}
\le
\sqrt{N- i_M}\sqrt{\sum_{i=i_M+1}^{N}s_{i,N}}\\
=&\sqrt{N- i_M}\sqrt{1- \sum_{i=1}^{i_M}s_i} \\
=&\sqrt{\Tr T - \Tr \left\{ \rho \ge \frac{1}{M}\right\}}
\sqrt{\Tr \rho \left\{ \rho \,< \frac{1}{M}\right\}}.
\end{align*}
Thus, we obtain (\ref{le3}).
\end{proof}

In order to treat PFLEC,
we have to consider
a measuring operation with LOCC.
Lo and Popescu characterize 
a projection valued measure $\{ P_{\omega}\}$ 
(Every $P_\omega$ is a projection, and 
$\sum_\omega P_\omega$ is the identity.) on the system $B$ as follows.
\begin{lem}\Label{lemLP1}{\rm\bf Lo and Popescu\cite{LP}}
For any projection valued measure $\{ P_{\omega,B}\}$ 
on the system $B$,
there exist a projection valued measure $\{ P_{\omega,A}\}$ 
on the system $A$
and local unitaries $U_{\omega,A}$ and $U_{\omega,B}$
such that
\begin{align}
(I \otimes P_{\omega,B})| \Phi\rangle
= (U_{\omega,A} \otimes U_{\omega,B})(P_{\omega,A} \otimes I )| \Phi\rangle.\Label{3-22-2}
\end{align}
\end{lem}
That is, if the initial pure state is known,
the operation corresponding to
any projection valued measurement on $B$
can be replaced by a projection valued measurement on $A$
and local unitaries on $A$ and $B$ based on measuring data. 
However, we have to treat a general 
measuring operation with LOCC.
The above Lo and Popescu's result can be 
generalized as follows.
\begin{lem}\Label{lemLP}
Given a measuring operation $I=\{I_\omega\}$ 
with LOCC on a tensor product space
${\cal H}_A \otimes {\cal H}_B$ 
and a pure state $| \Phi\rangle \langle \Phi|$
on the tensor product space ${\cal H}_A \otimes {\cal H}_B$,
there exist a POVM $\{ M_\omega \}$ 
(Every $M_\omega$ is a positive operator, and 
$\sum_\omega M_\omega$ is the identity.)
and the quantum operation $C_\omega$ 
with LOCC, such that
\begin{align}
I_\omega (\Phi)
=C_\omega( \sqrt{M_\omega} \otimes I | \Phi\rangle \langle \Phi|
\sqrt{M_\omega} \otimes I), \quad \forall \omega.
\end{align}
\end{lem}

\begin{proof}
It is known that 
any measuring operation $I_B=\{I_{\omega,B}\}$ on the system $B$
can be described by 
the projection valued measure $\{P_{\omega,B}\}$ on an extended space
${\cal H}_B'\supset {\cal H}_B$ and quantum operations 
$C_{\omega,B}$ on $B$ such that
\begin{align*}
I_{\omega,B}(\rho)= 
C_{\omega,B} (P_{\omega,B} \rho P_{\omega,B} ).
\end{align*}
Applying (\ref{3-22-2}),
we have
\begin{align*}
&(I_{\omega,B}\otimes I)(\Phi)\\
=& 
(I\otimes C_{\omega,B} )\Bigl(
(U_{\omega,A} \otimes U_{\omega,B})
(P_{\omega,A} \otimes I )
|\Phi\rangle \langle \Phi|\\
& \hspace{18ex} (P_{\omega,A} \otimes I )
(U_{\omega,A} \otimes U_{\omega,B})^*\Bigr).
\end{align*}
Hence, any operation on $B$ can be described by 
the combination of the projection measurement 
$\{P_{\omega,A}\}_\omega$ on $A$ and
local operations based only on the measuring data of 
$\{P_{\omega,A}\}_\omega$.

Now, we focus on 
a measurement operation $I'=\{I_\omega'\}_\omega$ on a tensor product space
${\cal H}_A \otimes {\cal H}_B$ consisting of LOCC
and a pure state $| \Phi\rangle \langle \Phi|$
on ${\cal H}_A \otimes {\cal H}_B$
satisfying 
the condition (A):
the set $\Omega= \{\omega\}$ consists of all sent 
classical informations.

Then, there exist the projection valued 
measure $\{P_{\omega,B}\}_\omega$ on an extended space
${\cal H}_B'\supset {\cal H}_B$ and quantum operations $C_{\omega,A}$ and $C_{\omega,B}$
such that
\begin{align*}
 I_{\omega}'(\Phi)
=
(C_{\omega,A}\otimes C_{\omega,B} )\Bigl(
(P_{\omega,A} \otimes I )
|\Phi\rangle \langle \Phi|
 (P_{\omega,A} \otimes I )
\Bigr).
\end{align*}
Even if the 
measurement operation $I=\{I_k\}_k$ 
with LOCC does not satisfies the condition (A),
there exist a measurement LOCC operation 
$I'=\{I_\omega'\}_{\omega\in \Omega}$ 
with subset $\Omega_k\subset \Omega$ 
satisfying the condition (A) such that
\begin{align}
I_k= \sum_{\omega \in \Omega_k}I_\omega'.
\end{align}
Hence, 
we have
\begin{align*}
& I_{k}(\Phi)\\
=&
\sum_{\omega \in \Omega_k}
(C_{\omega,A}\otimes C_{\omega,B} )\Bigl(
(P_{\omega,A} \otimes I )
|\Phi\rangle \langle \Phi|
 (P_{\omega,A} \otimes I )
\Bigr).
\end{align*}
That is,
there exist
a projection valued measure $\{\tilde{P}_{k,A}\}$ on 
an extended space ${\cal H}_A'\supset {\cal H}_A$
and LOCC operations
$C_k$ such that
\begin{align*}
 I_{k}(\Phi)
=
C_k
\bigl((\tilde{P}_{k,A} \otimes I )
|\Phi\rangle \langle \Phi|
 (\tilde{P}_{k,A} \otimes I )\Bigr).
\end{align*}
Since
the projection
$P_{{\cal H}_A}$ to ${\cal H}_A$
satisfies that 
$P_{{\cal H}_A}\tilde{P}_{k,A}P_{{\cal H}_A}
=(\tilde{P}_{k,A}P_{{\cal H}_A})^*
\tilde{P}_{k,A}P_{{\cal H}_A}$,
there exists a unitary $\tilde{U}_{k,A}$ such that
\begin{align*}
\tilde{P}_{k,A}P_{{\cal H}_A}
=
\sqrt{M_{k}^A}:= 
\tilde{U}_{k,A}
P_{{\cal H}_A}\tilde{P}_{k,A}P_{{\cal H}_A}.
\end{align*}
Hence, we obtain
\begin{align*}
& I_{k}(\Phi)\\
=&
C_k
\Bigl(
(\tilde{U}_{k,A}\otimes I )
(\sqrt{M_{k}^A} \otimes I )
|\Phi\rangle \langle \Phi|
 (\sqrt{M_{k}^A} \otimes I )
(\tilde{U}_{k,A}\otimes I )^*
\Bigr).
\end{align*}
Therefore, the proof is completed.
\end{proof}

In order to use the information spectrum method,
one may characterize the optimal failure probability
based on $\Tr\rho\{\rho-x\ge0\}$ for the reduced density $\rho$ 
of the initial state.
However, it is difficult. Hence, we focus on
$h(x):=\Tr(\rho-x)\{\rho-x\ge0\}$ instead of 
$\Tr\rho\{\rho-x\ge0\}$.
Suppose that we wish to
reduce all eigenvalues of the reduced density $\rho$ 
to be no greater than $x$.  This incurs a
probability of failure given by $h(x)$.
Upon success we obtain a normalized state whose
largest eigenvalue is not greater than $x/(1-h(x))$, 
which is majorized by a maximally entangled state of the dimension
$\lfloor (1-h(x))/x\rfloor $.  It turns out that this method is
optimal among PFLECs as follows.

\begin{figure}
\begin{center}
\scalebox{1.0}{\includegraphics[scale=0.40]{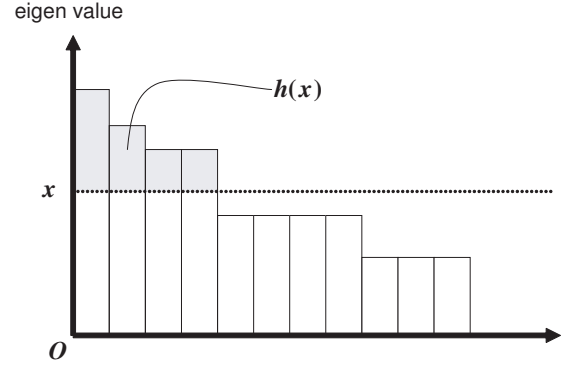}}
\end{center}
\caption{Illustration of $h(x)$}
\label{graph3}
\end{figure}%


\begin{lem}\Label{l2}
The bound on the performance of PFLEC based on
$\Phi$ is evaluated by using
the function $h(x)$, as follows:
\begin{align}
&\max_{I=\{I_0,I_1\}:\hbox{ PFLEC of } \Phi}
\{ L(I) | 
\Tr I_0(\Phi ) \le h(x)
\} \nonumber \\
&= \left\lfloor \frac{1}{x}( 1- h(x))\right\rfloor ,
\Label{le2}
\end{align}
where $\lfloor x \rfloor $ denotes the maximum integer $n$ satisfying
$n \le x$.
\end{lem}
\begin{proof}
From Lemma \ref{lemLP},
for any PFLEC $I$,
there exist two quantum operations $C_0$ and $C_1$ 
with LOCC 
and a positive operator $P$ such that
$0 \le P \le I $ and 
\begin{align}
\Tr I_0( \Phi )
&= \Tr (I-P) \rho \nonumber \\
\Tr I_1( \Phi )
&= \Tr P \rho \nonumber\\
I_1( \Phi )
&= C_1 ((\sqrt{I-P} \otimes I)|\Phi \rangle \langle \Phi|
(\sqrt{I-P} \otimes I))\nonumber\\
I_0( \Phi )
&= C_0 ((\sqrt{P} \otimes I)|\Phi \rangle \langle \Phi|
(\sqrt{P} \otimes I)).\nonumber
\end{align}
Hence, we obtain the following equations for the following reasons.
\begin{align}
& \min_{
I=\{I_0,I_1\}:\hbox{PFLEC of } \Phi}
\left\{\Tr I_0( \Phi )
\left | 
\frac{\Tr I_1( \Phi )}{L}
=x
\right.\right\}\nonumber \\
&=
\min_{I\ge P \ge 0 \hbox{ on } {\cal H}}
\{\Tr \rho(I-P)|
x - \sqrt{P}\rho\sqrt{P}  \ge 0\} \Label{7311} \\
&=
\min_{I\ge P \ge 0 \hbox{ on } {\cal H}}
\{\Tr (\rho-\sqrt{\rho}P\sqrt{\rho})|
x - \sqrt{\rho}P\sqrt{\rho}  \ge 0\} \Label{7314} \\
&= \min_{\sigma \hbox{ on } {\cal H}}
\{ 1- \Tr \sigma|
x - \sigma  \ge 0, \rho \ge \sigma \} \Label{7315} \\
&= \min_{\sigma \hbox{ on } {\cal H}}
\left\{\left. 1- \sum_i \langle e_i | \sigma | e_i \rangle  
\right|
\langle e_i | \sigma | e_i \rangle 
\le s_i, x \right\} \Label{7316} \\
&=
1- \sum_{i: s_i \le x} s_i - \sum_{i: s_i \,> x} x 
=
\Tr (\rho-x)\{\rho-x \ge 0\}
= h(x), \Label{7312}
\end{align}
where we diagonalize $\rho$ as $\rho= \sum_i s_i | 
e_i \rangle \langle e_i|$ in (\ref{7316}).
From Lemma \ref{l01},
there exists a quantum operation $C_1$ with LOCC
that transforms
the state $
\frac{1}{\Tr P \rho}
(P \otimes I)|\Phi \rangle \langle \Phi|
(P \otimes I)$
to a maximally entangled state with the size $L$
if and only if
$\frac{\Tr  I_1( \Phi )}{L}
\ge P \rho P$.
Thus, from Lemma \ref{lemLP}, we obtain (\ref{7311}).
In general, for any bounded operator $A$, there exists
a unitary operator $U$ such that
$A A^* = U A^* A U^*$.
Thus, the condition $x - P \rho P \ge 0 $ is
equivalent with the condition $x- \sqrt{\rho} P \sqrt{\rho} \ge 0$.
We obtain (\ref{7314}).
Replacing $\sqrt{\rho} P \sqrt{\rho}$ by $\sigma$,
we obtain (\ref{7315}).

Equation (\ref{7312}) implies
\begin{align*}
&\max_{I=\{I_0,I_1\}:\hbox{ PFLEC of } \Phi}
\left\{ L(I) \left| 
\Tr I_0( \Phi ) \le h(x)
\right.\right\}  \\
&= \max_{x'}\left\{ \left. \frac{1}{x'}( 1- h(x')) \right| 
\begin{array}{l}
\frac{1}{x'}( 1- h(x')) \hbox{ is an integer},\\
h(x') \le h(x) 
\end{array}\right\} \\
&= \left\lfloor \frac{1}{x}( 1- h(x))\right\rfloor ,
\end{align*}
where the second equation follows from the fact that the function $h(x)$
strictly monotonically decreases and is continuous.
\end{proof}
\section{Asymptotic theory}\Label{s5}
In this section, based on non-asymptotic formulas given in 
section \ref{s4},
we prove our main theorem.
For this purpose, we need to prepare 
the finite-version of 
the information-spectrum quantities 
for a projection operator $T_n$ 
and a reduced density $\sigma_n$ on ${\cal H}_{A,n}$ as follows.
\begin{align*}
\zeta_n (T_n| \sigma_n)&:= -\frac{1}{n}
\log \Tr \sigma_n T_n, \\
\zeta_{n,1/2} (T_n| \sigma_n)&:= -\frac{1}{n}
\log \Tr \sqrt{\sigma_n} T_n, \\
\eta_n  (T_n)&:= -\frac{1}{n}\log \Tr (I- T_n),\\
\zeta_n^c (T_n| \sigma_n)&:= -\frac{1}{n}
\log \Tr \sigma_n (I- T_n), \\
\zeta_{n,1/2}^c (T_n| \sigma_n)&:= -\frac{1}{n}
\log \Tr \sqrt{\sigma_n} (I- T_n).
\end{align*}
As the limiting version, we define
\begin{align*}
\overline{\zeta} (\vec{\bm{T}}| \vec{\bm{\sigma}})
&:=\varlimsup\zeta_n (T_n| \sigma_n),\\
\underline{\zeta} (\vec{\bm{T}}| \vec{\bm{\sigma}})
&:=\varliminf\zeta_n (T_n| \sigma_n),\\
\overline{\zeta}_{1/2} (\vec{\bm{T}}| \vec{\bm{\sigma}})
&:=\varlimsup\zeta_n (T_n| \sigma_n),\\
\underline{\zeta}_{1/2} (\vec{\bm{T}}| \vec{\bm{\sigma}})
&:=\varliminf\zeta_{n,1/2} (T_n| \sigma_n),\\
\overline{\eta} (\vec{\bm{T}}| \vec{\bm{\sigma}})
&:=\varlimsup\eta_n (T_n| \sigma_n),\\
\underline{\eta} (\vec{\bm{T}}| \vec{\bm{\sigma}})
&:=\varliminf\eta_n (T_n| \sigma_n),\\
\overline{\zeta}^c (\vec{\bm{T}}| \vec{\bm{\sigma}})
&:=\varlimsup\zeta_n^c (T_n| \sigma_n),\\
\underline{\zeta} ^c(\vec{\bm{T}}| \vec{\bm{\sigma}})
&:=\varliminf\zeta_n^c (T_n| \sigma_n),\\
\overline{\zeta}_{1/2}^c (\vec{\bm{T}}| \vec{\bm{\sigma}})
&:=\varlimsup\zeta_n^c (T_n| \sigma_n),\\
\underline{\zeta}^c_{1/2} (\vec{\bm{T}}| \vec{\bm{\sigma}})
&:=\varliminf\zeta_{n,1/2}^c (T_n| \sigma_n),
\end{align*}
for sequences $\vec{\bm{\sigma}}=\{\sigma_n\}$ and $\vec{\bm{T}}= \{T_n\}$.
For the projection $S_n(a):=
\{ \rho_n \,< e^{-na}\}$,
we simplify $\zeta_n (S_n(a)| \sigma_n),\zeta_{n,1/2} (S_n(a)| \sigma_n),$
$\eta_n  (S_n(a)), $ $\zeta_n^c (S_n(a)| \sigma_n)$, and
$\zeta_{n,1/2}^c (S_n(a)| \sigma_n)$
to $\zeta_n (a| \sigma_n),\zeta_{n,1/2} (a| \sigma_n),$
$\eta_n (a), \zeta_n^c (a| \sigma_n)$,
and $\zeta_{n,1/2}^c (a| \sigma_n)$.
We can similarly define
$\overline{\zeta} (a| \vec{\bm{\sigma}})$,
$\underline{\zeta}  (a| \vec{\bm{\sigma}})$,
$\overline{\zeta}_{1/2} (a| \vec{\bm{\sigma}})$,
$\underline{\zeta}_{1/2} (a| \vec{\bm{\sigma}})$,
$\overline{\eta} (a| \vec{\bm{\sigma}})$,
$\underline{\eta} (a| \vec{\bm{\sigma}})$,
$\overline{\zeta}^c (a| \vec{\bm{\sigma}})$,
$\underline{\zeta} ^c(a| \vec{\bm{\sigma}})$,
$\overline{\zeta}_{1/2}^c (a| \vec{\bm{\sigma}})$,
and $\underline{\zeta}^c_{1/2} (a| \vec{\bm{\sigma}})$.
Using these values, we can characterize 
the RHSs of (\ref{le1}), (\ref{le3}) and (\ref{le2}).
In particular, when a sequence 
$\vec{\bm{\sigma}}$ equals the sequence $\vec{\bm{\rho}}=\{ \rho_n\}$
of the reduced density of the given state,
we omit $\vec{\bm{\rho}}$ in the above values.

Moreover, to discuss the asymptotic theory,
we need to define the concept ``majorization'' 
in regard to sequences of reduced densities.
The sequence of reduced densities $\vec{\bm{\sigma}} =\{\sigma_n\}$ 
majorizes (strongly majorizes) 
another one $\vec{\bm{\sigma'}}=\{\sigma_n'\}$,
written $\vec{\bm{\sigma}} \succeq \vec{\bm{\sigma'}}$ 
($\vec{\bm{\sigma}} \succ \vec{\bm{\sigma'}}$)
if $\sigma_n \succeq \sigma_n'$ ($\sigma_n \succ \sigma_n'$),
respectively.

In the following, we proceed to the proof of our main theorem.
Before it, we should remark that in an asymptotic case,
we can neglect the gap between $\lfloor L_n\rfloor $ and $L_n$
because $L_n$ is large enough.
\begin{lem}\Label{l4}
Without any assumption, the equations 
\begin{align*}
B_{1}(\epsilon)=
B_{2}(\epsilon)=\sup_R\{ R | K(R) \le \epsilon\}.
\end{align*}
hold for every $\epsilon \in [0,1]$.
\end{lem}
\begin{proof}
From the definition,
the inequality
$B_{1}(\epsilon)\ge 
B_{2}(\epsilon)$
is trivial.
We only need to prove the two inequalities
\begin{align}
B_{2}(\epsilon) &\ge\sup_R\{ R | K(R) \le \epsilon\} 
\Label{le13}\\
B_{1}(\epsilon) &\le\sup_R\{ R | K(R) \le \epsilon\}.
\Label{le14}
\end{align}
Let $R$ be a real number satisfying
\begin{align}
K(R) \le \epsilon. \Label{le12}
\end{align}
From Lemma \ref{l2},
there exists a PFLEC $I^n$
such that
$\Tr I^n_0 (\Phi_n)
= h_n(e^{-n R})$
and $L_n = e^{n R}( 1- h_n(e^{-n R}))$,
where $h_n(x):= \Tr (\rho_n - x ) \{ \rho_n - x \ge 0\}$.
From (\ref{le12}), we have
\begin{align*}
 \lim \frac{1}{n} \log L_n =& R, \\
\varlimsup \Tr I^n_0 (\Phi_n)
\le &
\varlimsup \Tr \rho_n\{ \rho_n - e^{-n R} \ge 0\}\\
=& K(R) \le \epsilon.
\end{align*}
We have now obtained the direct part (\ref{le13}).

Next, we proceed to the converse part (\ref{le14}).
Let $I^n$ be a DFLEC satisfying
$\varliminf \langle\Psi_n|
C^n (\Phi_n)|\Psi_n \rangle
\ge 1- \epsilon$.
For any $R \,< \varliminf \frac{1}{n}\log L_n $,
we have 
\begin{align*}
\lim \frac{e^{n R}}{L_n} = 0.
\end{align*}
From Lemma \ref{l3}, for any $T_n$ satisfying
$\Tr T_n= L_n$, we have
\begin{align*}
& \frac{(\Tr \sqrt{\rho_n'}T_n)^2}{L_n} \\
\le &\frac{1}{L_n}
\Bigl( \sqrt{\Tr \{ \rho_n \ge e^{-n R}\}}
\sqrt{\Tr \rho_n \{ \rho_n \ge e^{-n R}\}}\\
&+\sqrt{L_n-\Tr \{ \rho_n \ge e^{-n R}\}}
\sqrt{\Tr \rho_n \{ \rho_n \,< e^{-n R}\}}\Bigr)^2
\\
= &
\Bigl( \sqrt{\frac{\Tr \{ \rho_n \ge e^{-n R}\}}
{L_n}}
\sqrt{\Tr \rho_n \{ \rho_n \ge e^{-n R}\}}\\
&+\sqrt{1-\frac{\Tr \{ \rho_n \ge e^{-n R}\}}
{L_n}}
\sqrt{\Tr \rho_n \{ \rho_n \,< e^{-n R}\}}\Bigr)^2 .
\end{align*}
Since $\lim 
\frac{\Tr \{ \rho_n \ge e^{-n R}\}}{L_n}
\le \lim \frac{e^{nR}}{L_n} = 0$,
\begin{align*}
1- \epsilon \le&
\varliminf
\frac{(\Tr \sqrt{\rho_n'}T_n)^2}{L_n}
\le
\varliminf
\Tr \rho_n \{ \rho_n \,< e^{-n R}\}\\
=& 1- K(R).
\end{align*}
Thus, we obtain (\ref{le14}).
\end{proof}

\begin{lem}\Label{l7}
We have
\begin{align*}
B_{e,D}(r) =B_{e,P}(r) =
\sup_R \{ R | \underline{\zeta}^c (R) \ge r \}.
\end{align*}
\end{lem}
\begin{proof}
Since
$B_{e,D}(r) \ge B_{e,P}(r)$,
we only need to prove the inequalities
\begin{align}
B_{e,D}(r)& \le \sup_R \{ R | \underline{\zeta}^c (R) \ge r \}
\Label{le21}\\
B_{e,P}(r) &\ge \sup_R \{ R | \underline{\zeta}^c (R) \ge r \}.
\Label{le22}
\end{align}
First, we prove the direct part (\ref{le22}).
Assume that $\underline{\zeta}^c(R) \ge r \,> 0$.
From Lemma \ref{l2},
for any $R$, there exists a PFLEC 
$I^n$
with the size $e^{n R}(1- (1- t_n(R))e^{-n\zeta^c_n(R)}))$
such that 
\begin{align*}
\Tr I_0^n(\Phi_n)
= (1- t_n(R))
e^{-n\zeta^c_n(R)}, 
\end{align*}
where 
\begin{align*}
t_n(R) := \frac{e^{-nR}\Tr \{\rho_n \ge e^{-nR}\}}
{\Tr \rho_n\{\rho_n \ge e^{-nR}\}}.
\end{align*}
Since $\underline{\zeta}^c(R) \,> 0$,
we have $0 \le (1- t_n(R))
e^{-n\zeta^c_n(R)} \le e^{-n\zeta^c_n(R)} \to 0$.
Thus,
we have the following relations
\begin{align*}
\lim \frac{1}{n} \log e^{n R}(1- (1- t_n(R))
e^{-n\zeta^c_n(R)})) &= R \\
\varliminf \frac{-1}{n}\log 
\Tr I_0^n(\Phi_n )
& \ge \underline{\zeta}^c(R) \ge r, 
\end{align*}
which imply the inequality (\ref{le22}).

Next, we proceed to the converse part (\ref{le21}).
Assume that the DFLEC $(C^n,\Psi_n)$
satisfies
\begin{align}
\varliminf \frac{1}{n} \log 
\left(1-
\langle \Psi_n | C^n (\Phi_n )| \Psi_n \rangle
\right)\ge r. \Label{28-1}
\end{align}
We define the projection $T_n$ and the reduced density $\rho_n'$
as
\begin{align*}
T_n := L_n \Tr_{{\cal H}_B} | \Psi_n \rangle \langle \Psi_n | ,
\quad 
\rho_n':= \argmax_{\rho'\succeq \rho_n}
\frac{\left(\Tr \sqrt{\rho'}T_n \right)^2}{L_n}.
\end{align*}
Then, Lemma \ref{l1} and (\ref{28-1}) yields that 
\begin{align*}
\varliminf \frac{1}{n} \log 
\left(1-
\frac{(\Tr \sqrt{\rho_n'}T_n)^2}{L_n}\right)
\ge r.
\end{align*}

For any $R' \,< R_0:= \varliminf \frac{1}{n} \log L_n $, there exists
an integer $N$ such that 
$R_n := \frac{1}{n} \log L_n \,> R'$ for $\forall n \ge N$.
When a projection $T_n$ satisfies that
$\Tr T_n = L_n$, Lemma \ref{l3} implies that
\begin{align}
& \frac{(\Tr \sqrt{\rho_n'}T_n)^2}{L_n}\nonumber \\
\le & \frac{1}{L_n}
\Bigl( \sqrt{\Tr \{ \rho_n \ge e^{nR'}\}}
\sqrt{\Tr \rho_n \{ \rho_n \ge e^{nR'}\}}\nonumber \\
&+ \sqrt{ L_n - \Tr \{ \rho_n \ge e^{nR'}\}}
\sqrt{\Tr \rho_n \{ \rho_n \,< e^{nR'}\}}
\Bigr)^2\nonumber \\
\le& \Bigl( 
e^{-\frac{n}{2}(\eta_n(R')+ \zeta^c_n(R')+R_n)}\nonumber \\
&+\sqrt{1- e^{-n(\eta_n(R')+R_n)}}
\sqrt{1-e^{-n \zeta^c_n(R')}}
\Bigr)^2\nonumber \\
 \cong &
 \Bigl( 1- \frac{1}{2} \left(
e^{-n(\eta_n(R')+R_n)} 
+ e^{-n\zeta^c_n(R')}
\right) \nonumber \\
&+ e^{-\frac{n}{2}(\eta_n(R')+ \zeta^c_n(R')+R_n)}
\Bigr)^2\nonumber \\
 =&
\left( 1- 
\frac{1}{2}
\left( 
e^{- \frac{n}{2}\zeta^c_n(R')}
-
e^{-\frac{n}{2}(\eta_n(R')+R_n)} 
\right)^2
\right)^2. \Label{ab1}
\end{align}
Since 
$e^{-\frac{n}{2}(\eta_n(R')+R_n)}
\le e^{-\frac{n}{2}(R_n- R')}
e^{-\frac{n}{2}(\eta_n(R')+R')}
\le e^{-\frac{n}{2}(R_n- R')}
e^{-\frac{n}{2} \zeta^c_n(R')}
\le e^{-\frac{n}{2} \zeta^c_n(R')}$, 
we have
\begin{align*}
& \left( 
e^{- \frac{n}{2}\zeta^c_n(R')}
-
e^{-\frac{n}{2}(\eta_n(R')+R_n)} 
\right)^2 \\
\ge&
(1- e^{-\frac{n}{2}(R_n- R')})^2
e^{-n \zeta^c_n(R')}.
\end{align*}
Thus,
\begin{align}
&\left( 1- 
\frac{1}{2}
\left( e^{-\frac{n}{2}(\eta_n(R')+R_n)} 
- e^{- \frac{n}{2}\zeta^c_n(R')}
\right)^2
\right)^2 \nonumber \\
\le &
\left( 1- 
\frac{1}{2}(1- e^{-\frac{n}{2}(R_n- R')})^2
e^{-n \zeta^c_n(R')}
\right)^2. \Label{ab2}
\end{align}
Since $\lim (1- e^{-\frac{n}{2}(R_n- R')})^2=1$,
it follows from (\ref{ab1}) and (\ref{ab2}) that
\begin{align*}
\underline{\zeta}^c(R')
\ge\varliminf \frac{1}{n} \log 
\left(1-
\frac{(\Tr \sqrt{\rho_n'}T_n)^2}{L_n}\right)
\ge r .
\end{align*}
Since $R'$ is an arbitrary real number satisfying $R' \,< R_0$,
the relation 
$R_0 \le \sup_R \{R | $ $\underline{\zeta}^c(R) \ge r \}$
holds.
Therefore, we obtain (\ref{le21}).
\end{proof}
\begin{lem}\Label{l6}
When $ \overline{\zeta} (a)=
\underline{\zeta}(a)=: \zeta(a)$
and there exists a real number $a$ such that 
$\zeta(a)\le \underline{\zeta}^c(a)$, 
\begin{align*}
&B_{e,P}^*(r)\\
=&
\sup_a \{
a-\min \{\zeta(a),a +\overline{\eta}(a)\} |
\min \{\zeta(a),a +\overline{\eta}(a)\} \le r\} \\
=&
\inf_a \{
a-\min \{\zeta(a),a +\overline{\eta}(a)\} |
\min \{\zeta(a),a +\overline{\eta}(a)\} \,> r\} \\
=&
\inf_a\{ a- \zeta (a) | \zeta(a) \le r \}.
\end{align*}
\end{lem}
\begin{proof}
First, we prove the direct part.
Consider a PFLEC $I^n$ satisfying
\begin{align*}
L_n &= \frac{1- h_n(e^{-na})}{e^{-na}}\\
\Tr I^n_0 ( \Phi_n )
&= h_n(e^{-na}).
\end{align*}
Thus, we have
\begin{align*}
\varliminf \frac{1}{n} \log L_n 
=&
a- \min \{\zeta(a),a +\overline{\eta}(a)\} ,\\
\varlimsup \frac{-1}{n} \log \left(
\Tr I^n_1 ( \Phi_n )\right)
=&\min \{\zeta(a),a +\overline{\eta}(a)\} .
\end{align*}
Therefore, we have
\begin{align*}
&B_{e,P}^*(r) \\
\ge&
\sup_a \{
a-\min \{\zeta(a),a +\overline{\eta}(a)\} |
\min \{\zeta(a),a +\overline{\eta}(a)\} \le r\} \\
=&
\max \left\{
\sup_a\{ a- \zeta(a) | \zeta(a)\le r \},
\sup_a \{ - \overline{\eta}(a)| a +\overline{\eta}(a)\le r\}
\right\}\\
=& \sup_a\{ a- \zeta(a) | \zeta(a)\le r \},
\end{align*}
where the final equation is derived by 
Lemma \ref{l61} as follows.
Using Lemma \ref{l61},
we have
$
\sup_a\{ a- \zeta(a) | \zeta(a)\le r \}
\ge
\sup_a \{ - \underline{\eta}(a)| a +\underline{\eta}(a)\le r\}
\ge
\sup_a \{ - \overline{\eta}(a)| a +\overline{\eta}(a)\le r\}$.

Next, we proceed to the converse part.
Let $\{ I^n\}$ be a sequence of PFLECs
such that $r \ge \varlimsup \frac{-1}{n}
\log (1-\epsilon_n )$, where
$\epsilon_n :=
\Tr I^n_0 ( \Phi_n )$.
In the following,
we focus on $\varliminf \frac{1}{n}\log L_n$.
Let $a$ be a real number satisfying 
\begin{align}
\varlimsup \frac{-1}{n}\log (1-\epsilon_n ) \le
r \le \min\{ \zeta(a), a + \overline{\eta}(a)\}.
\Label{le19}
\end{align}
Since 
\begin{align*}
& \varlimsup \frac{-1}{n} \log 
\left(\Tr \rho_n \{ \rho_n \le e^{-na}\} + e^{-na}\Tr
 \{ \rho_n \,> e^{-na}\}\right) \\
=&
\min\{ \zeta(a), a + \overline{\eta}(a)\},
\end{align*}
there exists an integer $N$ such that
\begin{align*}
&\Tr \rho_n \{ \rho_n \le e^{-na}\} + e^{-na}\Tr
 \{ \rho_n \,> e^{-na}\}\\
\ge &
1- h_n(e^{-na})
,\quad \forall n \ge N.
\end{align*}
Lemma \ref{l2} guarantees that
\begin{align}
&e^{na} \left(
\Tr \rho_n \{ \rho_n \le e^{-na}\} + e^{-na}\Tr
 \{ \rho_n \,> e^{-na}\}
\right)\nonumber \\
= &
\frac{1- h_n(e^{-na})}{e^{-na}}
\ge L_n. \Label{999}
\end{align}
Taking the limit of the exponent,
we have
\begin{align*}
\varliminf \frac{1}{n} \log L_n
\le a- \min \{ \zeta (a), a+ \overline{\eta}(a)\}.
\end{align*}
From (\ref{le19}),
we have 
\begin{align*}
&B_{e,P}^*(r)\\
\le &
\inf_a\{ a- \min \{ \zeta (a), a+ \overline{\eta}(a)\}
|
\min\{ \zeta(a), a + \overline{\eta}(a)\}
\ge r \}.
\end{align*}
It follows from (\ref{le17}) that
the function $a \mapsto 
\min \{ \zeta (a), a+ \overline{\eta}(a)\}$
is continuous.
Thus,
\begin{align*}
& \inf_a\{ a- \min \{ \zeta (a), a+ \overline{\eta}(a)\}
|
\min\{ \zeta(a), a + \overline{\eta}(a)\}
\ge r \}\\
&=
\sup_a \{a- \min \{ \zeta (a), a+ \overline{\eta}(a)\}
|\min\{ \zeta(a), a + \overline{\eta}(a)\}
\le r \}.
\end{align*}
The proof is now completed.
\end{proof}

\begin{lem}\Label{l5}
When $ \overline{\zeta} (a)=
\underline{\zeta}(a)=: \zeta(a)$
and there exists a real number $a$ such that 
$\zeta(a)\le \underline{\zeta}^c(a)$,
\begin{align}
&B_{e,D}^*(r)\nonumber \\
=&
\sup_{\vec{\bm{\rho'}}\succeq \vec{\bm{\rho}}}
\sup_{\vec{\bm{T}}}
\{
-\varliminf\eta_n(T_n) |
\varlimsup 2 \zeta_{n,1/2}^c(T_n|\rho'_n)- \eta(T_n)
\le r \} \Label{le39}\\
=&
\sup_{\vec{\bm{T}}}
\{
-\varliminf\eta_n(T_n) |
\varlimsup 2 \zeta_{n,1/2}^c(T_n)- \eta(T_n)
\le r \} \Label{le391}\\
=&
\sup_a \left\{ a -r \left| 
\inf_{a'} \left\{\zeta(a')- \frac{a'}{2}\right| 
a' \le a \right\}+ \frac{a}{2}
\le r \right\}  \Label{le38}\\
=&
\sup_a 
      \left.
           \left\{ 
                  \frac{a}{2} -
                   \inf_{a'} 
                  \left\{
                        \zeta(a')- \frac{a'}{2}
                  \right| 
                  a' \le a 
            \right\}
     \right|  \nonumber\\
&\hspace{13ex}
\inf_{a'} 
\left. \left\{
     \left.\zeta(a')- \frac{a'}{2}
     \right| 
     a' \le a 
\right\}+ \frac{a}{2}
\le r 
\right\} .\Label{le388}
\end{align}
\end{lem}
\begin{proof}
Equation (\ref{le39}) 
follows from (\ref{le1}).
Since the function
$a \mapsto 
\inf_{a'}\left\{
\left.\zeta(a')- \frac{a'}{2}\right| a' \le a \right\}$ is
continuous and decreases monotonically 
and
the function
$a \mapsto 
\inf_{a'}\left\{\left.
\zeta(a')- \frac{a'}{2}\right| a' \le a \right\}
$ $+ \frac{a}{2}$ is
continuous and
increases monotonically, 
equation (\ref{le388}) holds.
First, we prove the direct part:
\begin{align}
& \sup_{\vec{\bm{T}}}
\{
-\varliminf\eta_n(T_n) |
\varlimsup 2 \zeta_{n,1/2}^c(T_n)- \eta(T_n)
\le r \}  \nonumber \\
 \ge&
\sup_a 
\left.
\left\{ \frac{a}{2} -\inf_{a'} 
     \left\{
          \zeta(a')- \frac{a'}{2}
     \right| 
     a' \le a 
\right\}
\right| \nonumber \\
&\hspace{13ex}
\inf_{a'} 
\left.\left.\left\{
                  \zeta(a')- \frac{a'}{2}
            \right| 
            a' \le a 
      \right\}+ \frac{a}{2}
      \le r 
\right\} .
\Label{le392}
\end{align}
As we prove later, 
we can choose a projection $T_n(a,R)$
such that
\begin{align}
\eta_n(T_n(a,R))&= -R, \\
\zeta^c_{n,1/2}(T_n(a,R))&\le
\max\left\{ \zeta_{n,1/2}^c(a),
-R + \frac{a}{2} \right\}.\Label{le42}
\end{align}
When $\eta_n(a)\ge -R$,
the projection
$T_n(a,R):= \{ \rho_n - e^{-na}\ge 0\}$ satisfies 
(\ref{le42}).
Otherwise,
the projection $T_n(a,R):=
\{ \rho_n - e^{-na}\ge 0\} + 
(\{ \rho_n - e^{-na}\,< 0\}-\tilde{T}_n(a,R))$
satisfies (\ref{le42}),
where
$\tilde{T}_n(a,R)$
is constructed as follows:
We choose $m:=e^{nR}$
normalized eigenvectors $\{e_i'\}_{i=1}^m$ of 
$\{ \rho_n - e^{-na}\,< 0\}\rho_n$
in descending order concerning the eigenvalue,
and define the projection $\tilde{T}_n(a,R)$ by
$\sum_{i=1}^m | e_i' \rangle \langle e_i'|$.
The choice of $\{e_i'\}_{i=1}^m$ and 
the relation 
$e^{nR}=
\Tr \{ \rho_n- e^{-na}\,< 0\}e^{-n(-R-\eta_n(a))}$
guarantees
\begin{align}
\Tr \sqrt{\rho_n} \{ \rho_n - e^{-na}\,< 0\}  
e^{-n(-R-\eta_n(a))}
\le \Tr \sqrt{\rho_n} \tilde{T}_n(a,R) .
\end{align}
Then, we can check the condition (\ref{le42}) as follows:
\begin{align*}
&\eta_n(T_n(a,R))
= \frac{-1}{n}\log\Tr (I-T_n(a,R))\\
=&\frac{-1}{n}\log\Tr \tilde{T}_n(a,R)
= \frac{-1}{n}\log e^{nR} = -R,\\
&\zeta^c_{n,1/2}(T_n(a,R))
=\frac{-1}{n}\log\Tr\sqrt{\rho_n} \tilde{T}_n(a,R) \\
\le &
\frac{-1}{n}\log\Tr\sqrt{\rho_n}
\{ \rho_n - e^{-na}\,< 0\}  
e^{-n(-R-\eta_n(a))} \\
=& \zeta^c_{n,1/2}(a)-R -\eta_n(a)
\le -R +\frac{a}{2}.
\end{align*}
Now, we apply Lemma \ref{l51}
to the case $\rho_n=\rho_n, \sigma_n = \sqrt{\rho_n}$.
Since 
\begin{align*}
&\{\rho_n -e^{na}\sigma_n > 0\}
=\{\rho_n -e^{na}\sqrt{\rho_n} > 0\}\\
=&\{\sqrt{\rho_n} -e^{na} > 0\}
=\{\rho_n -e^{2na} > 0\},
\end{align*}
we have
\begin{align}
\underline{\eta}(a)= \underline{\zeta}_{1/2}^c(2a|\vec{\bm{\rho}}), \quad
\underline{\zeta}(a)= \underline{\zeta}(2a|\vec{\bm{\rho}}). \Label{28-2}
\end{align}
From Lemma \ref{l51},
the maximum $a_r$ of 
\begin{align*}
\left\{
a\left|\inf_{a'} \left\{\zeta(a')- \frac{a'}{2}\right| 
a' \le a \right\}+ \frac{a}{2}
= r \right\} 
\end{align*}
exists.
We define $R$ by
\begin{align*}
R:= \frac{a_r}{2}-\inf_{a'}\left\{ \left.
\zeta(a')- \frac{a'}{2}\right|
a' \le a_r\right\} .
\end{align*}
Then, $R$ equals to the right hand side of (\ref{le392}),
and we have
\begin{align*}
&\varlimsup 2 \zeta_{n,1/2}^c(T_n(a_k,R))- \eta(T_n(a_k,R))\\
\le &
2 \max \left\{
\overline{\zeta}_{1/2}(a_r+1/k), -R+ \frac{a_r+1/k}{2}
\right\} \\
\le & r+ 1/k,
\end{align*}
where $a_k := a_r + 1/k$ and $k$ is a fixed integer,
and the last inequality follows from 
(\ref{ap11}) in Lemma \ref{l51} in Appendix \ref{as1}.
We define $N_k$ as the minimum integer satisfying
\begin{align*}
2 \zeta_{n,1/2}^c(T_n(a_k,R))- \eta(T_n(a_k,R))
\le r + \frac{2}{k},
\quad \forall n \ge N_k.
\end{align*}
For the sequence $b_n:=
\min_k \{ a_k | n \ge N_k\}$,
we have 
\begin{align}
\varlimsup 2 \zeta_{n,1/2}^c(T_n(b_n,R))- \eta(T_n(b_n,R))
\le r. \Label{b7}
\end{align}
Inequality (\ref{le392}) follows from 
(\ref{b7}) and 
the first equation of (\ref{le42}).

Next, we prove the converse part.
Assume that $\{ (T_n, \rho_n')\}$ satisfies 
$\varlimsup_{n\to \infty}
 2 \zeta_{n,1/2}^c(T_n|\rho'_n)- \eta(T_n)\le r$.
There exists a subsequence $\{n_k\}$
such that $\lim \eta_{n_k} (T_{n_k})= -R_0
:= \varliminf \eta_n (T_n)$.
Focusing on the projection
$\{ \rho_n' -e^{-na}\ge 0\}=\{\sqrt{\rho_n'}-e^{-na/2} \ge 0\}$, 
we have
\begin{align*}
&\Tr \sqrt{\rho_n'}
\{ \rho_n' -e^{-na}\ge 0\}
- e^{na/2}\Tr \{ \rho_n' -e^{-na}\ge 0\} \\
\ge &
\Tr \sqrt{\rho_n'} (I-T_n)
- e^{na/2}\Tr (I-T_n) ,
\end{align*}
which implies 
\begin{align*}
&\Tr \sqrt{\rho_n'}
\{ \rho_n' -e^{-na}\ge 0\}
+ e^{na/2}\Tr (I-T_n)\\
\ge & \Tr \sqrt{\rho_n'} (I-T_n).
\end{align*}
Taking the limit $k\to \infty$,
we have
\begin{align*}
\min\left\{
\varlimsup_{k \to \infty}
\zeta^c_{n_k,1/2}(a| \rho_{n_k}')
,\frac{a}{2}- R_0 \right\}
\le
\varlimsup_{k \to \infty}
\zeta^c_{n_k,1/2}(T_{n_k}| \rho_{n_k}').
\end{align*}
Now, we apply Lemma \ref{l9} to the case $\rho_n= \rho_n',
\sigma= \sqrt{\rho_n'}$.
In this case, similarly to (\ref{28-2}),
we have
\begin{align*}
\underline{\eta}(a)= \underline{\zeta}_{1/2}^c(2a|\vec{\bm{\rho'}}), \quad
\underline{\zeta}(a)= \underline{\zeta}(2a|\vec{\bm{\rho'}}). 
\end{align*}
Hence, (\ref{le30}) yields that
\begin{align*}
& \varlimsup_{k \to \infty}\zeta^c_{n_k,1/2}(a| \rho_{n_k}')
\ge \underline{\zeta}^c_{1/2}(a|\vec{\bm{\rho'}}) \\
\ge &\inf_{a'}
\left\{ \left. \underline{\zeta}(a'|\vec{\bm{\rho'}})- \frac{a'}{2}
\right| a' \le a \right\}.
\end{align*}
Since $\rho_n' \succeq \rho_n$, we have
$\underline{\zeta}(a'|\vec{\bm{\rho'}}) \ge
\underline{\zeta}(a')= \zeta(a')$, {\it i.e.,}
\begin{align*}
\inf_{a'}
\left\{ \left. \underline{\zeta}(a'|\vec{\bm{\rho'}})- \frac{a'}{2}
\right| a' \le a \right\}
\ge \inf_{a'}
\left\{ \left. \zeta(a')- \frac{a'}{2}
\right| a' \le a \right\}.
\end{align*}
Thus,
\begin{align}
r \ge & \varlimsup_{n\to \infty}
 2 \zeta_{n,1/2}^c(T_n|\rho'_n)- \eta(T_n) \nonumber\\
\ge &
\varlimsup_{k \to \infty}
2 \zeta_{n_k,1/2}^c(T_{n_k}|\rho'_{n_k})- 
\eta(T_{n_k})\nonumber \\
\ge &
2 \min\left\{
\varlimsup_{k \to \infty}
\zeta^c_{n_k,1/2}(a| \rho_{n_k}')
,\frac{a}{2}- R_0 \right\}
+ R_0  \nonumber \\
\ge&
2 \min\left\{
\inf
\left\{ \left. \zeta(a')- \frac{a'}{2}
\right| a' \le a \right\}
,\frac{a}{2}- R_0 \right\}
+ R_0  \Label{le45}.
\end{align}
Since the function $a \mapsto 
\frac{a}{2} -\inf_{a'}
\left\{ \left. \zeta(a')- \frac{a'}{2}
\right| a' \le a \right\}$ is continuous,
there exists a real number $a$ 
such that
\begin{align*}
R_0
= \frac{a}{2} -\inf_{a'}
\left\{ \left. \zeta(a')- \frac{a'}{2}
\right| a' \le a \right\}.
\end{align*}
Using (\ref{le45}), we have
\begin{align*}
r \ge \inf_{a'}
\left\{ \left. \zeta(a')- \frac{a'}{2}
\right| a' \le a \right\}+ \frac{a}{2},
\end{align*}
which implies 
\begin{align*}
R_0 
\le &
\sup_a
\left\{
\frac{a}{2} -\inf_{a'}
\left\{ \left. \zeta(a')- \frac{a'}{2}
\right| a' \le a \right\}
\right|\\
&\hspace{14ex} \left.\inf_{a'}
\left\{ \left. \zeta(a')- \frac{a'}{2}
\right| a' \le a \right\}+ \frac{a}{2}
\le r \right\}.
\end{align*}
The proof is now completed.
\end{proof}

\section{Relation to random number generation}\Label{s7}
As a related problem,
it is known to transform from a given known probability distribution 
$p$ to
a desired probability distribution $q$.
If it is possible, 
the majorization relation 
$q \succeq p$ holds.
However, even if
the majorization relation 
$q \succeq p$ holds,
this transformation is not necessarily available.
Hence, if the two entangled pure states $\Phi_1$ and $\Phi_2$ 
have Schmidt coefficients corresponding to $p$ and $q$,
the Quantum LOCC operation transforming
from $\Phi_1$ to $\Phi_2$ 
is easier than
transform from $p$ to $q$.

In particular, when the desired distribution is 
the uniform distribution,
this problem is called
intrinsic randomness.
In this problem, 
our operation of intrinsic randomness
is described by
the map $\psi$ from the original space $\Omega$
to ${\cal M}=\{1, \ldots, M\}$.
When the initial distribution is $p$
and the uniform distribution is described by $p_M$
on ${\cal M}$,
one of criteria of its quality is 
the half of the square of Hellinger distance between 
$p\circ \psi^{-1}$ and $p_M$:
\begin{align}
\varepsilon(\psi,p)
:=
1- \sum_{i=1}^M
\sqrt{
\frac{\sum_{\omega \in \psi^{-1}(i)}p_\omega}
{M}}.
\end{align}
In this case, we describe the size of 
its target uniform distribution
$\psi$
by $M(\psi)$.
Hence, for a sequence of the initial distributions $\{p_n\}$,
we can define
the optimal rates 
\begin{align*}
B_H(\epsilon)&:=
\sup_{\{\psi_n\}} 
\Bigl\{\varliminf \frac{\log M(\psi_n) }{n}
\Bigl|
\varlimsup \varepsilon(\psi_n,p_n)
\le \epsilon \Bigr\} \\
B_{e,H}(r)&:=
\sup_{\{\psi_n\}} 
\Bigl\{
\varliminf \frac{\log M(\psi_n)}{n}
\Bigr|
\varliminf \frac{-1}{n}\log \varepsilon(\psi_n,p_n)
\ge r
\Bigr\} \\
B_{e,H}^*(r)&:=
\sup_{\{\psi_n\}} \Bigl\{
\varliminf \frac{\log M(\psi_n)}{n}
\Bigr|\\
&\hspace{18ex}
\varlimsup \frac{-1}{n}\log (1-\varepsilon(\psi_n,p_n))
\le r
\Bigr\} .
\end{align*}
The variational distance version with 
the constant constraint
has been discussed by Vembu \& Verd\'{u} \cite{V-V} and Han \cite{Han_book}.

Let $\Phi_n$ be the entangled pure state with the Schmidt coefficient 
corresponding to $p_n$.
When $C_n$ is the quantum LOCC operation corresponding to
$\psi_n$
and $\Psi_n$ is the maximally entangled state with the size $M(\psi_n)$,
we have
\begin{align}
1- \varepsilon(\psi_n,p_n)
= \sqrt{\langle \Psi_n |C_n (\Phi_n)|\Psi_n\rangle}, \Label{29-1}
\end{align}
{\it i.e.},
\begin{align}
2 \varepsilon(\psi_n,p_n)- \varepsilon(\psi_n,p_n)^2
= 1- \langle \Psi_n |C_n (\Phi_n)|\Psi_n\rangle. \Label{29-2}
\end{align}
Hence, 
comparing the entanglement concentration with 
the initial entangled state $\Phi_n$
and the intrinsic randomness with the 
initial distribution $p_n$,
(\ref{29-2}) yields that 
\begin{align*}
B_H(\epsilon) \le
B_D(2\epsilon-\epsilon^2).
\end{align*}
Since 
\begin{align*}
\varepsilon(\psi_n,p_n)
\le 1- \langle \Psi_n |C_n (\Phi_n)|\Psi_n\rangle
\le 2 \varepsilon(\psi_n,p_n),
\end{align*}
the inequality
\begin{align*}
B_{e,H}(r) \le
B_{e,D}(r)
\end{align*}
holds.
Moreover, the equation (\ref{29-1}) yields that
\begin{align*}
B_{e,H}^*(r) \le
B_{e,D}^*(2r).
\end{align*}

When we adopt
the KL divergence criterion:
\begin{align*}
D(p_M\|p\circ \psi^{-1})
:= \log M + \sum_{i=1}^M
\frac{1}{M}\log 
\left(\sum_{\omega \in \psi^{-1}(i)}p_\omega\right),
\end{align*}
we focus on
the following value:
\begin{align*}
B_{KL}(\epsilon)&:=
\sup_{\{\psi_n\}} 
\Bigl\{\varliminf \frac{\log M(\psi_n) }{n}
\Bigl|
\varlimsup D(p_{M(\psi_n)}\|p\circ \psi_n^{-1})
\le \epsilon \Bigr\} .
\end{align*}
As is shown Hayashi\cite{H-second},
the relation
\begin{align*}
B_{KL}(\epsilon)=
\sup_a \{ a - \zeta(a)| \zeta(a) < \epsilon\}
\end{align*}
holds.
When $\zeta(a)$ is continuous,
\begin{align}
B_{KL}(\epsilon)=
B_{e,P}^*(\epsilon).
\end{align}
In particular, if the limit of R\'{e}nyi entropy is differentiable,
\begin{align}
B_{KL}(\epsilon)\ge
B_{e,H}^*(\epsilon/2)
\end{align}
when $\epsilon \le -\frac{1}{2}\overline{\psi}'\left(\frac{1}{2}\right)
-\overline{\psi}\left(\frac{1}{2}\right)$.
The above relation is an interesting relation between 
Hellinger criterion and 
KL divergence criterion.

\section{Concluding remarks}
We derive asymptotic bounds based on several formulations from 
Lemma \ref{l1}, \ref{l3}, and \ref{l2}.
Since these bounds are tight in a general source,
the evaluations given in Lemma \ref{l1}, 
\ref{l3}, and \ref{l2}
are useful 
in a non asymptotic case
as well as
in an asymptotic case.
Even if the class of DFLEC is wider than
that of PFLEC,
their asymptotic performances
are almost equivalent.
A difference appears only between $B_{e,D}^*(r)$
and $B_{e,P}^*(r)$.
For example, 
when the limit of R\'{e}nyi entropy $\overline{\psi}(s)$ is differentiable,
$B_{e,D}^*(r)$ is larger than  $B_{e,P}^*(r)$
if and only if $r$ is greater than 
$- \frac{1}{2}\overline{\psi}'\left(\frac{1}{2}\right)- 
\overline{\psi}\left(\frac{1}{2}\right)$.
From (\ref{le391}) of Lemma \ref{l5},
the bound $B_{e,D}^*(r)$ can be attained without an LOCC,
{\it i.e.}, the original reduced density $\rho_n$ is close enough to
an appropriate MES only in regard to $B_{e,D}^*(r)$.
As a byproduct, in Appendix \ref{as1},
we establish several general relations between 
information-spectrum quantities.
\appendix
\subsection{General relations for information
spectrums}\Label{as1}
Here, we prove some lemmas required by our proof. 
In this section,
we treat information-spectrum quantities with 
more general definitions, which are given in Nagaoka and Hayashi\cite{NH}.
This is because we need such a general treatment in our proof of 
Lemma \ref{l5}.

For the two sequences $\{\rho_n\}$ and $\{\sigma_n\}$ of trace class
positive semidefinite operators,
we discuss how to characterize an information-spectrum quantity
$\underline{\eta}(a)
:= \varliminf \frac{-1}{n} \log \Tr \sigma_n \{
\rho_n - e^{-na} \sigma_n \,> 0 \}$
by using two other information-spectrum quantities $\underline{\zeta}(a)
:= \varliminf \frac{-1}{n} $ $\log \Tr \rho_n \{
\rho_n - e^{-na} \sigma_n \le 0 \}$ and $
\underline{\zeta}^c(a)
:= \varliminf \frac{-1}{n} \log \Tr \rho_n \{
\rho_n - e^{-na} \sigma_n \,> 0 \}$.
As discussed later, when 
$\overline{\zeta}(a)
:= $ $\varlimsup \frac{-1}{n} \log $ $\Tr \rho_n \{
\rho_n - e^{-na} \sigma_n \le 0 \}$
equals $\underline{\zeta}(a)$ for any $a$,
we can use the same method to characterize 
another information spectrum $\overline{\eta}(a)
:= \varlimsup \frac{-1}{n} \log \Tr \sigma_n \{
\rho_n - e^{-na} \sigma_n \,> 0 \}$.
As was proven by Nagaoka and Hayashi\cite{NH},
the function $\underline{\zeta}(a)$ increases
monotonically, and other functions $
\underline{\zeta}^c(a)$ and $\underline{\eta}(a)$
decrease monotonically \cite{NH}.
Focusing on the projection
$\{\rho_n-e^{-na}\sigma_n \ge 0\}$,
we have
\begin{align*}
\Tr (\rho_n - e^{-na}\sigma_n) \{\rho_n-e^{-na}\sigma_n\ge 0\} \ge 0,
\end{align*}
which yields to 
\begin{align*}
\Tr \rho_n \{\rho_n-e^{-na}\sigma_n\ge 0\} \ge
e^{-na}\Tr \sigma_n \{\rho_n-e^{-na}\sigma_n\ge 0\} .
\end{align*}
Thus, we have
\begin{align}
\underline{\zeta}^c(a)
\le \underline{\eta}(a) + a. \Label{le31}
\end{align}
Similarly,
we can prove
\begin{align*}
&\Tr (\rho_n - e^{-na}\sigma_n) \{\rho_n-e^{-na}\sigma_n\ge 0\} \\
\ge &
\Tr (\rho_n - e^{-na}\sigma_n) \{\rho_n-e^{-nb}\sigma_n\ge 0\} .
\end{align*}
By adding $e^{-na} \Tr \sigma_n$ to both sides,
we have
\begin{align*}
&\Tr \rho_n \{\rho_n-e^{-na}\sigma_n\ge 0\}
+  e^{-na}\Tr \sigma_n \{\rho_n-e^{-na}\sigma_n\,< 0\} \\
& \ge
\Tr \rho_n \{\rho_n-e^{-nb}\sigma_n\ge 0\}
+  e^{-na}\Tr \sigma_n \{\rho_n-e^{-nb}\sigma_n\,< 0\}.
\end{align*}
Taking the limit $n \to \infty$, we obtain
\begin{align}
\min \{ \underline{\zeta}(a),a+\underline{\eta}(a)\}
\ge 
\min \{ \underline{\zeta}(b),a+\underline{\eta}(b)\} 
\Label{le17}
\end{align}
for any $a$ and $b$\cite{NH}.
When $\underline{\zeta}(a)=\overline{\zeta}(a)$ for any $a$,
we can replace $\underline{\eta}$
by $\overline{\eta}$.
From inequality (\ref{le17}), We can derive the following two formulas;
\begin{align}
\underline{\eta}(a)+a \ge \underline{\zeta}(b)
&\hbox{ if }
\underline{\eta}(b) \,>
\underline{\eta}(a) 
\Label{c3} \\
\underline{\zeta}(a)  \ge a+ \underline{\eta}(b)
&\hbox{ if }
\underline{\zeta}(a) \,<
\underline{\zeta}(b) \Label{c4},
\end{align}
which play important roles in the following lemmas.
As a lower bound of $\underline{\eta}(a)$,
the following lemma holds.
\begin{lem}\Label{l9}
If there exists a real number $a_0$ such that
$\underline{\zeta}(a_0) \le \underline{\zeta}^c(a_0)$,
the relations
\begin{align}
\underline{\eta}(a) 
&\ge \inf _{a'}\{ \underline{\zeta}(a') - a' |
a' \,< a \}  \Label{le32}\\
&= \inf_{a'} \{ \underline{\zeta}(a') - a' |
a' \le a \} \Label{le30}
\end{align}
hold.
\end{lem}
\begin{proof}
From (\ref{le31}),
the relations
\begin{align*}
\underline{\zeta}(a_0) \le \underline{\zeta}^c(a_0)
\le \underline{\eta}(a_0) + a_0
\end{align*}
hold.
Since $\underline{\eta}(a_0)\ge 
\underline{\zeta}(a_0)- a_0$,
we have 
\begin{align*}
\underline{\eta}(a_0)
\ge\inf_{a'} \{ \underline{\zeta}(a') - a' |
a' \le a_0 \}.
\end{align*}
For any $a \le a_0 $, 
the relation 
$\underline{\zeta}(a) \le \underline{\zeta}^c(a)$
holds.
Since $\underline{\zeta}(a-0) 
\le \underline{\zeta}(a)$,
the equation (\ref{le30}) holds.
Similarly, we can prove that
a real number $a(\le a_0)$ satisfies (\ref{le32}).

Next, we prove (\ref{le32})
for any $a \,> a_0$
by the transfinite induction.
Assume that the relation (\ref{le32}) holds
for any real number $b$ satisfying
$a \,> b$ and
\begin{align}
\underline{\eta}(a)
\,< \inf_{a'} \{ \underline{\zeta}(a') - a' |
a' \le a \}. \Label{le341}
\end{align}
For any $\epsilon \,> 0$,
we have
\begin{align*}
\underline{\eta}(a)
\,< \inf_{a'} \{ \underline{\zeta}(a') - a' |
a' \le a - \epsilon \}
\le \underline{\eta}(a- \epsilon).
\end{align*}
From (\ref{c3}),
we have $
\underline{\eta}(a) \ge \underline{\zeta}(a-\epsilon) 
-a$.
Since $\epsilon$ is arbitrary,
we obtain the inequality
\begin{align*}
\underline{\eta}(a)
\ge
\inf_{a'} \{ \underline{\zeta}(a') - a' |
a' \,< a \},
\end{align*}
which contradicts assumption (\ref{le341}).
\end{proof}

The following lemma is another characterization
of the lower bounds of $\underline{\eta}(a)$.
\begin{lem}\Label{l61}
We obtain the inequality
\begin{align*}
\sup_a\{a- \underline{\zeta}(a) | \underline{\zeta}(a)\le r \}
\ge \sup_a \{
- \underline{\eta}(a) |
a+ \underline{\eta}(a)\le r \},
\end{align*}
which is equivalent to
another inequality
\begin{align*}
\inf_a\{ \underline{\zeta}(a) -a | \underline{\zeta}(a)\le r \}
\le \inf_a \{
\underline{\eta}(a) |
a+ \underline{\eta}(a)\le r \}.
\end{align*}
\end{lem}
\begin{proof}
We prove it by reduction to absurdity.
Assume that there exists a real number $a_0$ such that
\begin{align}
a_0 + \underline{\eta}(a_0) & \le r , \Label{la1}\\
- \underline{\eta}(a_0) & \,>
\sup_a\{a- \underline{\zeta}(a) | \underline{\zeta}(a)\le r \}.\Label{la2}
\end{align}
We will lead contradiction with the two cases,
case 1: 
$a_1 := \inf_a \{ a |\underline{\eta}(a)=
\underline{\eta}(a_0)\}> a_0$,
case 2: $a_1=a_0$.

In case 1, for any real number $\epsilon \in (0,a_0-a_1)$,
the inequality $\underline{\eta}(a_1- \epsilon)
\,> \underline{\eta}(a_1 + \epsilon)
$ holds. Using (\ref{c3}), we have
\begin{align*}
& \underline{\zeta}(a_1- \epsilon)
\le
\underline{\eta}(a_1+ \epsilon) + a_1 + \epsilon
=
\underline{\eta}(a_0) + a_1 +\epsilon \\
\le &
r +(a_1-a_0) + \epsilon 
< r .
\end{align*}
Thus,
\begin{align*}
& \sup_a\{a- \underline{\zeta}(a) | \underline{\zeta}(a)\le r \}
\ge 
a_1-\epsilon - \underline{\zeta}(a_1- \epsilon) \\
\ge &
a_1-\epsilon - (a_1+\epsilon)
- \underline{\eta}(a_1+ \epsilon) 
=  - \underline{\eta}(a_0)- 2\epsilon.
\end{align*}
Taking the limit $\epsilon  \to 0$,
we obtain 
$\sup\{a- \underline{\zeta}(a) | \underline{\zeta}(a)\le r \}\ge
- \underline{\eta}(a_0) $,
which contradicts (\ref{la2}).

In case 2, the inequality $\eta(a_0) \,<  
\underline{\eta}(a_0 - \epsilon)$
holds for $\forall \epsilon \,> 0$.
Using (\ref{c3}),
we have $\underline{\zeta}(a_0 - \epsilon) \le 
\underline{\eta}(a_0) + a_0 \le r$.
Thus,
\begin{align*}
& \sup_a\{a- \underline{\zeta}(a) | \underline{\zeta}(a)\le r \}
\ge
a_0 - \epsilon - \underline{\zeta}(a_0 - \epsilon)\\
\ge & a_0 - \epsilon - a_0 - \underline{\eta}(a_0)
= -\epsilon - \underline{\eta}(a_0).
\end{align*}
This also contradicts (\ref{la2}).
\end{proof}

Define the sets
$I$ and $I'$ as
\begin{align*}
I&:=
\{ a \in \real | \underline{\zeta}(a)
\,> \underline{\zeta}(a- \epsilon) 
\quad \forall \epsilon \,> 0 \}, \\
I'&:=
\{ a \in \real | \underline{\zeta}(a+ \epsilon)
\,> \underline{\zeta}(a) 
\quad \forall \epsilon \,> 0 \} .
\end{align*}
As upper bounds of $\underline{\eta}(a)$,
we have the following two lemmas.
\begin{lem}\Label{l10}
We have two inequalities
\begin{align}
\underline{\eta}(a) 
&\le \inf_{a \in I} \{ \underline{\zeta}(a') - a' |
a' \le a \}  \Label{le35},\\
\underline{\eta}(a) 
&\le \inf_{a \in I'} \{ \underline{\zeta}(a') - a' |
a' \,< a \} \Label{le34}.
\end{align}
If $\underline{\zeta}(a) =
\overline{\zeta}(a) $ for any real $a$,
we have two other inequalities
\begin{align}
\overline{\eta}(a) 
&\le \inf_{a \in I} \{ \underline{\zeta}(a') - a' |
a' \le a \}  \Label{le36},\\
\overline{\eta}(a) 
&\le \inf_{a \in I'} \{ \underline{\zeta}(a') - a' |
a' \,< a \} \Label{le37}.
\end{align}
\end{lem}
\begin{proof}
First, we prove (\ref{le35}).
Let $a'\in I$ be a real number
satisfying $a' \le a$.
From (\ref{c4}), we have 
\begin{align*}
a'-\epsilon + \underline{\eta}(a')
\le \underline{\zeta}(a'-\epsilon) , \quad
\forall \epsilon \,> 0 .
\end{align*}
Since $\epsilon \,>0$ is arbitrary,
we obtain the relation 
\begin{align*}
\underline{\eta}(a)\le
\underline{\eta}(a')
\le \underline{\zeta}(a'-0)-a'
\le \underline{\zeta}(a')-a'.
\end{align*}
From the arbitrariness of $a'$, 
the above relation implies (\ref{le35}).
Similarly, we can prove (\ref{le36}).

Next, we prove (\ref{le34}).
Let $a'\in I'$ be a real number
satisfying $a' \,< a$.
From (\ref{c4}), we have 
\begin{align*}
a' + \underline{\eta}(a'+\epsilon)
\le \underline{\zeta}(a').
\end{align*}
If $\epsilon \,> 0$ is small enough,
\begin{align*}
\underline{\eta}(a) 
\le 
\underline{\eta}(a'+\epsilon)
\le \underline{\zeta}(a') - a'.
\end{align*}
From the arbitrariness of $a'$, 
the above inequality implies (\ref{le34}).
Similarly, we can prove (\ref{le37}).
\end{proof}
\begin{lem}\Label{l51}
Assume that a real number $r$ satisfies that
\begin{align}
r \,< \sup_a 
\left\{\inf_{a'}\left\{\left.\underline{\zeta}(a')- a'\right| 
a' \le a \right\}+ a
\right\}. \Label{ap13}
\end{align}
The maximum $a_r$ of 
\begin{align}
\left\{
a\left|\inf_{a'} \left\{\underline{\zeta}(a')- a'\right| 
a' \le a \right\}+ a
= r \right\}\Label{ap99}
\end{align}
exists.
Moreover, the inequality
\begin{align}
\underline{\eta}(a_r +\epsilon) \le
\inf_{a} \left.\left\{\underline{\zeta}(a)- a\right| 
a \le a_r \right\} ,\quad
\forall \epsilon \,> 0 \Label{ap11}
\end{align}
holds.
When $ \overline{\zeta} (a)=
\underline{\zeta}(a)$ for any $a$,
we can replace $\underline{\eta}$ by $\overline{\eta}$ in the above argument.
\end{lem}
\begin{proof}
Since the function
$g:a \mapsto 
\inf_{a'} \left\{\left.\underline{\zeta}(a')- a'\right| 
a' \le a \right\}+ a$ is continuous
and increases monotonically,
it follows from (\ref{ap13})
that set (\ref{ap99}) is bounded and closed.
Thus the maximum of the set (\ref{ap99}) exists.

Next, we prove (\ref{ap11}).
First we assume that
\begin{align}
\underline{\zeta}(a_r) - a_r  \ge
\inf_{a'} \left\{\left.\underline{\zeta}(a')- a'\right| 
a' \le a_r \right\},\Label{ap10c}
\end{align}
Since the function $g$ increases monotonically
and $a_r+\epsilon$ does not belong to the set (\ref{ap99}),
the relations
\begin{align*}
& \underline{\zeta}(a) <
\underline{\zeta}(a_r) 
=
\inf_{a'} \left\{\left.\underline{\zeta}(a')- a'\right| 
a' \le a_r \right\}+ a_r =r\\
 \,< &
\inf_{a'} \left\{\left.\underline{\zeta}(a')- a'\right| 
a' \le a_r + \epsilon \right\}
+ a_r+ \epsilon  \le\underline{\zeta}(a_r +\epsilon)
\end{align*}
hold for $a<a_r$.
Applying (\ref{c4}) to the case $b=a_r+\epsilon$,
we obtain (\ref{ap11}).

Second, we assume the opposite inequality
\begin{align}
\underline{\zeta}(a_r) - a_r  \,< 
\inf_{a'} \left\{\left.\underline{\zeta}(a')- a'\right| 
a' \le a_r \right\}.\Label{ap10}
\end{align}
There exists a sequence $\{ a_n \}$ such that
\begin{align*}
&\underline{\zeta}(a_n) - a_n \to 
\inf_{a'} \left\{\left.\underline{\zeta}(a')- a'\right| 
a' \le a_r \right\} \\
&a_n \,< a_r.
\end{align*}
From the above relations,
there exists an integer $N$ such that
$\underline{\zeta}(a_n) \,< \underline{\zeta}(a_r), \quad \forall n \ge N$.
Using (\ref{c4}), we have
\begin{align*}
\underline{\eta}(a_r) \le 
\underline{\zeta}(a_n)- a_n.
\end{align*}
Thus, we obtain
\begin{align}
\underline{\eta}(a_r ) \le
\inf_{a'} \left\{\left.\underline{\zeta}(a')- a'\right| 
a' \le a_r \right\} \Label{ap12},
\end{align}
which implies (\ref{ap11}).
\end{proof}

\subsection{G\"{a}rtner-Ellis theorem}\Label{GE}
Here, for our proof of Theorem \ref{t-g},
we discuss G\"{a}rtner-Ellis theorem \cite{DZ}.
Let $X_n$ be a sequence of random variables.
Then, the logarithmic moment function is defined as
\begin{align*}
\Lambda_n(t):= \log {\rm E}_{X_n} e^{t X_n},
\end{align*}
where ${\rm E}_{X_n}$ denotes the expectation concerning the 
random variable $X_n$.
The logarithmic moment function $\Lambda_n(t)$ is convex.
\begin{thm}
Assume that
the limit $\Lambda(t):= \lim_{n \to \infty}\frac{\Lambda_n(t)}{n}$
exists.
Then, defining
the rate function
\begin{align}
\Lambda^*(R):=\sup_{t}t R-\Lambda(t),
\end{align}
we have
\begin{align}
\varliminf \frac{-1}{n}\log {\rm P}_{X_n}
\left\{\frac{X_n}{n} \ge a\right\}
&\ge  \inf_{R \ge a}\Lambda^*(R) \\
\varlimsup \frac{-1}{n}\log {\rm P}_{X_n}
\left\{\frac{X_n}{n} > a\right\}
&\le  \inf_{R > a}\Lambda^*(R) \\
\varliminf \frac{-1}{n}\log {\rm P}_{X_n}
\left\{\frac{X_n}{n} \le a\right\}
&\ge  \inf_{R \le a}\Lambda^*(R) \\
\varlimsup \frac{-1}{n}\log {\rm P}_{X_n}
\left\{\frac{X_n}{n} < a\right\}
&\le  \inf_{R < a}\Lambda^*(R).
\end{align}
\end{thm}
Using the above theorem, we can show the following theorem.
Since the function $\Lambda_n(t)$ is convex,
the $\Lambda(t)$ is convex, too.
Hence, when we choose the real numbers $R_1, R_2, R_3$ and $R_4$ as
\begin{align}
R_1&:= \lim_{t \to \infty}\frac{\Lambda(t)}{t}, \quad
R_2:= \lim_{t \to +0}\frac{\Lambda(t)}{t}, \\
R_3&:= \lim_{t \to -0}\frac{\Lambda(t)}{t}, \quad
R_4:= \lim_{t \to -\infty}\frac{\Lambda(t)}{t},
\end{align}
the relations
\begin{align}
R_4 \le R_3\le R_2 \le R_1
\end{align}
hold.
Thus, as is proven latter, the equations
\begin{align}
&\lim \frac{-1}{n}\log {\rm P}_{X_n}
\left\{\frac{X_n}{n} \ge a\right\}=
\lim \frac{-1}{n}\log {\rm P}_{X_n}
\left\{\frac{X_n}{n} > a\right\}\nonumber \\
=&\left\{
\begin{array}{ll}
0 & \hbox{if }a \le R_2 \\
\displaystyle \max_{t > 0} t R -\Lambda(t) > 0 & \hbox{if } R_2 < a < R_1\\
\infty & \hbox{if } R_1 < a 
\end{array}
\right. \Label{28-3}
\end{align}
and
\begin{align}
&\lim \frac{-1}{n}\log {\rm P}_{X_n}
\left\{\frac{X_n}{n} \le a\right\}=
\lim \frac{-1}{n}\log {\rm P}_{X_n}
\left\{\frac{X_n}{n} < a\right\}\nonumber\\
=&\left\{
\begin{array}{ll}
0 & \hbox{if } R_3 \le a\\
\displaystyle\max_{t < 0} t R -\Lambda(t) > 0 & \hbox{if } R_4 < a < R_3\\
\infty & \hbox{if } a < R_4
\end{array}
\right.\Label{28-4}
\end{align}
hold.
Moreover,
if the function $\Lambda$ is differentiable
at $t_0 >0$,
and if $R_2 < a < \Lambda'(t_0)$,
we have
\begin{align}
&\lim \frac{-1}{n}\log {\rm P}_{X_n}
\left\{\frac{X_n}{n} \ge a\right\}=
\lim \frac{-1}{n}\log {\rm P}_{X_n}
\left\{\frac{X_n}{n} > a\right\}\nonumber \\
=&
\sup_{t_0 \ge t > 0} t R -\Lambda(t)  .\Label{28-5}
\end{align}

\par\noindent\quad{\it Proof of (\ref{28-3}), (\ref{28-4}) 
and (\ref{28-5}):}\quad
First, we calculated the rate function $\Lambda^*(a)$.
When $R_3 \le a \le R_2$,
\begin{align*}
\Lambda^*(a)=\sup_{t} t a -\Lambda(t) = 0 a -\Lambda(0)=0.
\end{align*}
Assume that $R_2 < a < R_1$.
Then, if $\epsilon>0$ is sufficiently small,
\begin{align*}
&\Lambda^*(a)=\sup_{t} t a -\Lambda(t) 
\ge
\epsilon a -\Lambda(\epsilon ) = 
(a-R_2)\epsilon  + R_2 \epsilon  - \Lambda(\epsilon )\\
&\cong 
(a-R_2)\epsilon  + R_2 \epsilon  
- \lim_{t \to +0}\frac{\Lambda(t)}{t} \epsilon  
= (a-R_2)\epsilon  > 0.
\end{align*}
Now, we choose 
$t_{a}\neq 0$ such that
$t_{a} a =\Lambda(t_{a})$.
The convexity of $\Lambda$ guarantees that
\begin{align*}
\Lambda^*(a)=\sup_{t} t a -\Lambda(t) = 
\max_{0\le t\le t_{a}} t a -\Lambda(t).
\end{align*}
For $a$ such that $R_2\le a' \le a$,
since $t_a \ge t_{a'}$,
we have
\begin{align*}
\Lambda^*(a')=
\max_{0\le t\le t_{a'}} t a' -\Lambda(t)
=
\max_{0\le t\le t_{a}} t a' -\Lambda(t).
\end{align*}
Hence, the function $\Lambda^*$ is continuous $[R_2,a]$.
Thus, the function $\Lambda^*$ is continuous $[R_2,R_1)$.
in addition, when $a > R_1$,
$\Lambda^*(a)=\infty$.
Hence, 
when $a < R_1$,
we obtain
\begin{align*}
&\inf_{R \ge a}\Lambda^*(R) 
=\inf_{R > a}\Lambda^*(R) 
\nonumber \\
=&\left\{
\begin{array}{ll}
0 & \hbox{if }a \le R_2 \\
\displaystyle \max_{t > 0} t R -\Lambda(t) > 0 & \hbox{if } R_2 < a < R_1\\
\end{array}
\right.
\end{align*}
When $a > R_1$,
\begin{align*}
\inf_{R \ge a}\Lambda^*(R) 
=\inf_{R > a}\Lambda^*(R) 
= \infty.
\end{align*}
Therefore, we obtain (\ref{28-3}).
Similarly, we can prove (\ref{28-4}).

Moreover, 
for $a$ such that $R_2 \le a < R_1$,
we choose $t'_{a}=\argmax_{t}t a -\Lambda(t)$.
The convexity of $\Lambda$ guarantees that
when $R_2 \le a' <a$,
we have $t'_{a'}\le t'_{a}$.
Therefore, we prove (\ref{28-5}).
\endproof

Finally, in order prove (\ref{66-a}) and (\ref{66})
in our proof of Theorem \ref{t-g},
we focus on 
the probability distributions $p_n = \{p_{n,i}\}$,
and apply the above discussion to 
the random variable $- \log p_{n,i}$.
Using (\ref{28-3}), (\ref{28-4}) 
and (\ref{28-5}), we obtain (\ref{66-a}) and (\ref{66}).


\begin{thebibliography}{99}
\bibitem{H-second}
M. Hayashi,
``Second order asymptotics in fixed-length source coding and
intrinsic randomness,''
e-print: cs.IT/0503089.
\end{thebibliography}
\end{document}